\documentclass[rmp,aps,showpacs,preprintnumbers,amsmath,amssymb,twocolumn,superscriptaddress,longbibliography]{revtex4-2}
\usepackage{graphicx}
\usepackage{dcolumn}
\usepackage{bm}
\usepackage{amssymb}
\usepackage{amsmath}
\usepackage{epstopdf}
\usepackage{booktabs}
\usepackage{threeparttable}
\usepackage[pdfstartview=FitH, CJKbookmarks=true, bookmarksnumbered=true, bookmarksopen=true, colorlinks, pdfborder=001, linkcolor=blue, anchorcolor=blue, citecolor=blue]{hyperref}

\usepackage{soul,color}
\setstcolor{red}

\begin{document}
\title{Unconventional fully-gapped superconductivity in the heavy-fermion metal CeCu$_2$Si$_2$
}

\author{Michael Smidman}
\email{msmidman@zju.edu.cn}
\affiliation{Center for Correlated Matter and School of Physics, Zhejiang University, Hangzhou 310058, China}
\author{Oliver Stockert}
\email{oliver.stockert@cpfs.mpg.de}
\affiliation{Max Planck Institute for Chemical Physics of Solids, 01187 Dresden, Germany}
\author{Emilian M. Nica}
\affiliation{Department of Physics, Arizona State University, Tempe, AZ 85281, USA}
\author{Yang Liu}
\affiliation{Center for Correlated Matter and School of Physics, Zhejiang University, Hangzhou 310058, China}
\author{Huiqiu Yuan}
\email{hqyuan@zju.edu.cn}
\affiliation{Center for Correlated Matter and School of Physics, Zhejiang University, Hangzhou 310058, China}
\affiliation  {State Key Laboratory of Silicon Materials, Zhejiang University, Hangzhou 310058, China}
\author{Qimiao Si}
\email{qmsi@rice.edu}
\affiliation{Department of Physics and Astronomy, Rice Center for Quantum Materials, Rice University, Houston, TX 77005, USA}
\author{Frank Steglich}
\email{Frank.Steglich@cpfs.mpg.de}
\affiliation{Center for Correlated Matter and School of Physics, Zhejiang University, Hangzhou 310058, China}
\affiliation{Max Planck Institute for Chemical Physics of Solids, 01187 Dresden, Germany}
	
	\date{\today}
	
	\addcontentsline{toc}{chapter}{Abstract}

  \begin{abstract}

The heavy-fermion metal CeCu$_2$Si$_2$ was the first discovered unconventional, non-phonon-mediated superconductor, and  for a long time was believed to exhibit single-band $d$-wave superconductivity, as inferred from various measurements hinting at a nodal gap structure. More recently however, measurements using a range of techniques  at very low temperatures ($T \lessapprox 0.1$~K) provided evidence for a fully-gapped superconducting order parameter. In this Colloquium, after a brief historical overview we survey the apparently conflicting results of numerous experimental studies on this compound. We then address the different theoretical scenarios which have been applied to understand the  particular gap structure, including both isotropic (sign-preserving) and anisotropic two-band $s$-wave superconductivity, as well as an effective 
two-band $d$-wave model, where the latter can explain the currently available experimental data on CeCu$_2$Si$_2$. The lessons from CeCu$_2$Si$_2$ are expected to help uncover the Cooper-pair states in other unconventional, fully-gapped superconductors with strongly correlated carriers, and in particular highlight the rich variety of such states enabled by orbital degrees of freedom.

	\end{abstract}
	
	\maketitle
\tableofcontents
	\section{Introduction}		

Strongly correlated electron systems are central to contemporary studies of quantum materials. In these materials, electron-electron interactions have a strength that reaches or even exceeds the width of the underlying noninteracting electron bands. This property is to be contrasted with conventional metals such as aluminum or ordinary semiconductors like silicon, where electronic properties can be successfully described in terms of  noninteracting electrons with a materials-specific bandstructure. Instead, for strongly correlated electron systems, the interactions lead to rich emergent phenomena and novel electronic phases of matter. Examples of strongly correlated electron systems include cuprate perovskites \cite{Lee2006,Proust2019}, iron-based pnictides and chalcogenides \cite{Si2016,Stewart2011}, organic charge-transfer salts \cite{Maple2004,Lang2003,Kanoda2008},  and the moir\'{e} structures of graphene and transition-metal dichalcogenides \cite{Cao2018,Andrei2020}.

Among the strongly correlated electron systems, heavy fermion compounds such as CeCu$_2$Si$_2$ take a special place. The reason is simple. These materials contain partially-filled $f$-orbitals. For these $f$-electrons, the interactions are larger than their bandwidth to such an extent that the $f$-electrons act as localized magnetic moments. Indeed, at the heart of the physics of heavy fermion materials is the Kondo effect, whereby localized magnetic moments situated in a sea of conduction electrons become screened, and below a characteristic temperature scale (the Kondo temperature $T_K$), the local moments are entirely quenched leaving behind a remanent nonmagnetic Kondo singlet \cite{Hewson1997}. Such screened moments act as strong elastic scatterers, accounting for the peculiar logarithmic increase of the resistivity upon cooling when small concentrations of  certain magnetic impurities are introduced into nonmagnetic metals \cite{Kondo1964}. As was detected by \onlinecite{Triplett1971} for the dilute magnetic alloys \textbf{Cu}Cr and \textbf{Cu}Fe, the impurity-derived ``incremental'' low-temperature specific heat is proportional to temperature, $\Delta C(T) = \gamma T$, with a huge coefficient $\gamma$, that exceeds the Sommerfeld coefficient of the host metal Cu by more than a factor of 1000. This indicates the formation of a narrow local Kondo resonance at the Fermi level $E_F$ and could be well described in the framework of a local Fermi liquid theory \cite{Nozieres1974}.

Heavy fermion metals comprise of two broad classes,  lanthanides and actinides.  The lanthanide-based variants are commonly considered to be ideal examples of Kondo lattice systems. These materials rather than having a dilute random distribution of  local moments instead host a dense, periodic lattice of Kondo ions \cite{Aliev1983b,Brandt1984,Fulde1988,Kuramoto2012,Ott1987,Stewart1984a}. The first observation of heavy-fermion phenomena, i.e., the properties of a heavy Fermi liquid, was reported for the hexagonal paramagnetic compound  CeAl$_3$ \cite{Andres1975}. Here, the low-temperature specific heat, which is practically identical to the $4f$-electron contribution, was found to be proportional to temperature with a $\gamma$ coefficient of the same gigantic size as the aforementioned value for \textbf{Cu}Fe. In addition, the low-temperature resistivity of CeAl$_3$ was observed to follow a $\Delta \rho(T) = AT^2$ dependence with a huge prefactor $A$. These early findings were ascribed to a $4f$ virtual bound state at $E_F$. A very similar large $\gamma$ coefficient of the low-$T$ specific heat to that of CeAl$_3$ could be estimated for the putative paramagnetic phase of the  cubic antiferromagnet CeAl$_2$ (with a similar $T_K$) by treating the Ce ions as isolated Kondo centers \cite{Schotte1975}. This was taken as striking evidence for the heavy fermion phenomena in these Ce compounds indeed being due to the many-body Kondo effect rather than one-particle physics \cite{Bredl1978}. 

\begin{figure}[t]
	\begin{center}
		\includegraphics[width=0.65\columnwidth]{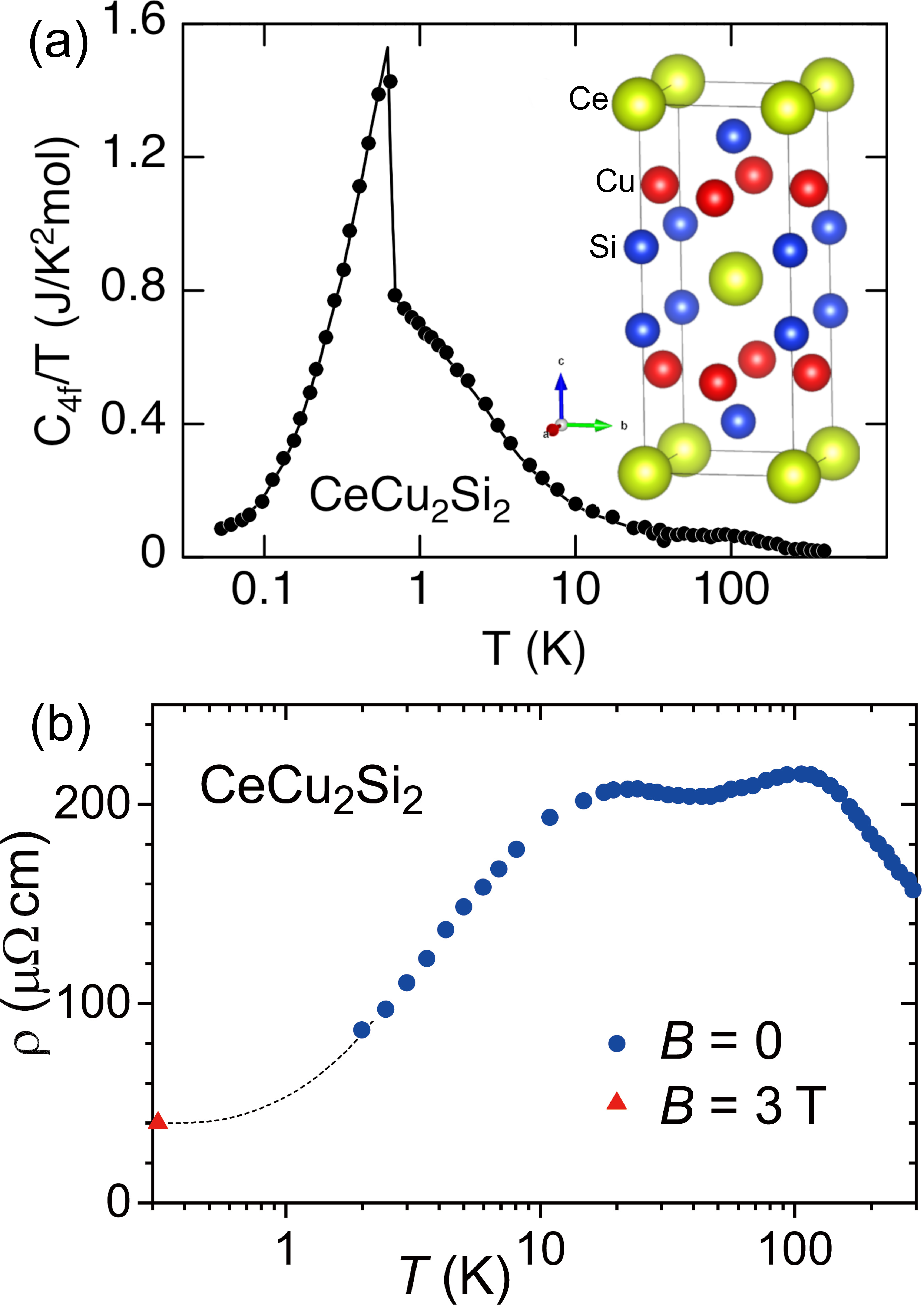}
	\end{center}
	\caption{(a) Contribution of the $4f$-electrons to the specific heat of CeCu$_2$Si$_2$, plotted as $C_{4f}/T$ vs. $T$ on a logarithmic scale. Replotted from \onlinecite{Steglich1990}. The solid line is a guide to the eyes. The inset shows the crystal structure of CeCu$_2$Si$_2$ (ThCr$_2$Si$_2$-type structure, space group $I4/mmm$), where green, red, and blue spheres correspond to Ce, Cu and Si atoms (see labels), respectively. (b) Temperature dependence of the resistivity of CeCu$_2$Si$_2$ on a logarithmic temperature scale, reproduced from \onlinecite{Shan2022}.}
	\label{Fig1new}
\end{figure}

The participation of the $f$-electrons in the electronic structure at sufficiently low temperatures causes the renormalized electronic bands to take on significant ‘$f$-electron’ character and the effective mass of the charge carriers exceeds that of ordinary conduction electrons by a factor up to about a thousand \cite{Zwicknagl1992}. This leads to the aforementioned unusual behaviors of canonical heavy-fermion compounds such as CeCu$_2$Si$_2$, namely the  $\gamma$ coefficient is of the order of J/K$^2$mol (Fig.~\ref{Fig1new}(a)), and there is a correspondingly enhanced temperature-independent Pauli spin susceptibility \cite{Sales1976,Grewe1991} (Fig.~\ref{Fig1}). As displayed in Fig.~\ref{Fig1new}(b), the electrical resistivity first exhibits an increase upon cooling from high temperatures, reflecting increasing incoherent scattering similar to that which occurs from dilute magnetic impurities. At lower temperatures however, Kondo lattice effects set in whereby coherent scattering of conduction electrons from the Kondo singlets below a characteristic temperature ($T_K\approx 15$~K for CeCu$_2$Si$_2$  \cite{Stockert2011}) leads to a pronounced decrease of the resistivity \cite{Coleman2007}. In several heavy-fermion metals, this decline of the resistivity follows a Fermi-liquid-type $AT^2$ dependence with a huge $A$ coefficient, whereas CeCu$_2$Si$_2$ exhibits non-Fermi-liquid behavior, as discussed  in Sec.~\ref{QCPSec}.

Another stark difference between Kondo lattices and the dilute impurity case is that in the former the Kondo effect  competes with the indirect Ruderman, Kittel, Kasuya and Yoshida (RKKY) magnetic exchange interaction \cite{Ruderman1954,Kasuya1956,Yosida1957} which tends to stabilize the $f$-electron moments. While predominant Kondo screening results in a paramagnetic heavy-fermion ground state, a dominant RKKY interaction causes magnetic, most frequently antiferromagnetic order. For quite a substantial number of these heavy-fermion metals the Kondo screening turns out to almost exactly cancel the RKKY interaction in the zero-temperature limit, which may give rise to a continuous zero-temperature quantum phase transition or quantum critical point (QCP), that can be easily accessed by adjusting a suitable non-thermal control parameter, e.g., pressure, doping or magnetic fields \cite{Stewart2001,Gegenwart2008,Si2010,Sachdev2011}. To get rid of the large residual entropy accumulated at the QCP, symmetry-broken novel phases are often observed, notably ‘unconventional’ superconductivity which cannot be accounted for by the electron-phonon mediated pairing mechanism of Bardeen-Cooper-Schrieffer (BCS) theory \cite{Norman2011,Norman2013,Stewart2017}.

The heavy fermion metal CeCu$_2$Si$_2$ was also the first  unconventional superconductor to be discovered \cite{Steglich1979} (Table~\ref{table:table1}), and it has recently attracted much research interest again. While it was considered a (single-band) $d$-wave superconductor for many years \cite{Ishida1999,Fujiwara2008}, the observation of a fully developed energy gap at very low temperatures \cite{Kittaka2014,Yamashita2017,Takenaka2017,Pang2018} has led to proposals of CeCu$_2$Si$_2$ being a two-band $s$-wave superconductor both with \cite{Ikeda2015,Li2018}, and without \cite{Yamashita2017,Takenaka2017,Tazai2018,Tazai2019} a sign-change of the order parameter. 

In this Colloquium article, after a brief historical overview we discuss the seemingly conflicting results of a large number of experimental studies on this material and address the different theoretical models applied to understand the 
particular gap structure. These models are divided into two categories. One class builds on a normal state in the presence of Kondo-driven renormalization, and utilizes the multiplicity of orbitals to realize a new kind of pairing state. In the band basis, this takes the form of  a band-mixing $d+d$-pairing state \cite{Nica2021}, in parallel with the proposed pairing state for the iron chalcogenides that are among the highest-$T_{\rm c}$ Fe-based superconductors \cite{Nica2017} based on strongly orbital-selective electron correlations. The other class directly works in the band basis, treats the Coulomb repulsive interaction perturbatively, and constructs a pairing state using the standard procedure of 
finding irreducible representations of the crystalline lattice's point group. This is exemplified by the 
$s_{+-}$ scenario  \cite{Ikeda2015,Li2018}, by analogy to a similar construction  applied to the Fe-based superconductors \cite{Mazin2008} in which a repulsive interband interaction leads to different signs of the order parameter between hole and electron pockets. We summarize the details of these considerations throughout the body of the article. 
In addition, we 
suggest that the insights gained from the analysis of the pairing state in CeCu$_2$Si$_2$ will have
broad implications on strongly correlated superconductivity in multi-orbital systems, and discuss future efforts 
that may shed further light on this canonical problem in the field of strongly correlated electron systems.

\section{The history of heavy-fermion superconductivity}

\begin{figure}[t]
	\begin{center}
		\includegraphics[width=0.99\columnwidth]{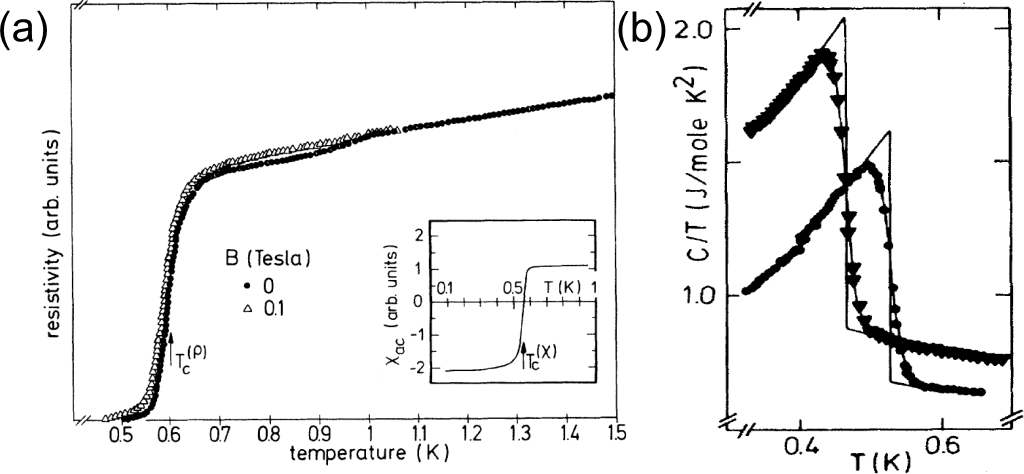}
	\end{center}
	\caption{(a) Resistivity $\rho(T)$, ac-susceptibility  $\chi_{\rm ac}(T)$ and (b) specific heat as $C/T$ vs $T$ for polycrystalline CeCu$_2$Si$_2$ indicating bulk superconductivity at $T_{\rm c}\approx0.5$~K, after \onlinecite{Steglich1979}. The Pauli susceptibility ($T>T_{\rm c}$) shown in the inset amounts to $\chi_P = 82\times10^{-9}{\rm m}^3$/mol \cite{Aarts1984}. Note that the normal-state values of both $\rho(T)$ and $C(T)/T$ point to non-Fermi-liquid behavior. In panel (b) data of two samples are displayed which have the same nominal composition and were prepared in the same way. Reproduced with permission from \onlinecite{Steglich1979}.}
	\label{Fig1}
\end{figure}

Given the strong pair-breaking effect of diluted localized spins in conventional superconductors \cite{Matthias1958,Abrikosov1960}, the discovery of bulk superconductivity in CeCu$_2$Si$_2$ \cite{Steglich1979} came as a big surprise. In a BCS superconductor, a tiny amount of randomly distributed magnetic impurities fully suppresses the superconducting state \cite{Maple1968,Riblet1971,Maple1972,Steglich1974}, but the superconductivity is robust against doping with nonmagnetic impurities \cite{Anderson1959,Balatsky2006}. On the other hand, superconductivity in CeCu$_2$Si$_2$ relies on a periodic array of 100 at\% magnetic Ce$^{3+}$ ions, each containing a localized $4f$-shell occupied by one electron in a  $J=5/2$ Hund’s rule ground state. Figure~\ref{Fig1} displays the first reported evidence for the superconducting transition in CeCu$_2$Si$_2$ at $T_{\rm c}\approx0.5$~K on annealed polycrystalline samples. Upon cooling through $T_{\rm c}$, the electrical resistivity falls to zero from a normal state with a non-saturated, nearly linear temperature dependence, while the ac susceptibility undergoes a rapid change from a strongly enhanced Pauli paramagnetic susceptibility to a large diamagnetic value [Fig.~\ref{Fig1}(a)].

Two early observations have led to the conclusion that CeCu$_2$Si$_2$ must be an unconventional bulk superconductor: (i) the nonmagnetic reference compound LaCu$_2$Si$_2$ is not a superconductor, at least down to 20 mK \cite{Steglich1979}, and (ii) a  tiny amount of nonmagnetic (as well as magnetic) substitution at the level of 1 at \% may lead to a complete suppression of superconductivity in CeCu$_2$Si$_2$ \cite{Spille1983}, see Sec.~\ref{impSec}. Further evidence for this conclusion could be drawn from  the specific-heat results shown in Fig.~\ref{Fig1}(b). Here the normal-state values of $C(T)/T$ are of the order of several hundreds of  mJ/mol~K$^2$; they substantially increase upon lowering the temperature and extrapolate to about 1~J/mol~K$^2$ in the zero-temperature limit. This exceeds the Sommerfeld coefficient of the electronic specific heat of Cu by more than a factor of 1000 and this proves that similar to CeAl$_3$, the measured specific heat in this low-temperature range is practically identical with the electronic contribution ($\approx C_{4f}$). The corresponding renormalized kinetic energy, $k_{\rm B}T_{\rm F}^{*}$, corresponds to the Kondo screening energy, $k_{\rm B}T_{\rm K}$, $T_{\rm K}\approx$~15~K \cite{Stockert2011}. Therefore, the ratio $T_{\rm c}/T_{\rm F}^{*}$ is of the order of 0.04, compared to $T_{\rm c}/T_{\rm F}\approx10^{-3}-10^{-4}$ for an ordinary BCS superconductor, highlighting CeCu$_2$Si$_2$ as a ‘high-$T_{\rm c}$ superconductor’ in a normalized sense \cite{Steglich1979}. On the other hand, the ratio $T_{\rm F}^{*}/\theta_{\rm D}$, where $\theta_{\rm D}$ is the Debye temperature, also amounts to about 0.05, while in a main  group metal or transition metal, $T_{\rm F}/\theta_{\rm D}$ is of order 100. The latter warrants the electron-phonon coupling in conventional BCS superconductors to be retarded, such that the Coulomb repulsion between conduction electrons is minimized and isotropic $s$-wave Cooper pairs may be formed.

For heavy-fermion metals, such phonon-mediated on-site pairing is clearly prohibited because of their low renormalized Fermi velocity which is, at best, of the order of the velocity of sound. Nevertheless, an early proposal was put forward to explain heavy-fermion superconductivity in CeCu$_2$Si$_2$ by a coupling of the heavy charge carriers to the breathing mode \cite{Razafimandimby1984}, while recently such a phonon-mediated superconductivity for this compound was expected to be realized near a magnetic instability, thanks to the vertex corrections due to  multipole charge fluctuations \cite{Tazai2018}. On the other hand, a broad consensus evolved shortly after the discovery of heavy-fermion superconductivity that here an electronic pairing mechanism must be operating \cite{Machida1983,Tachiki1984}. Therefore, CeCu$_2$Si$_2$ was soon regarded generally as an unconventional, i.e., non-phonon-driven, superconductor. Because of the phenomenological similarity of  heavy-fermion superconductivity in CeCu$_2$Si$_2$ with the superfluidity in $^3$He \cite{Osheroff1972a,Osheroff1972b}, a magnetic coupling mechanism appeared to be most natural \cite{Anderson1984}.

\begin{figure}[t]
	\begin{center}
		\includegraphics[width=0.9\columnwidth]{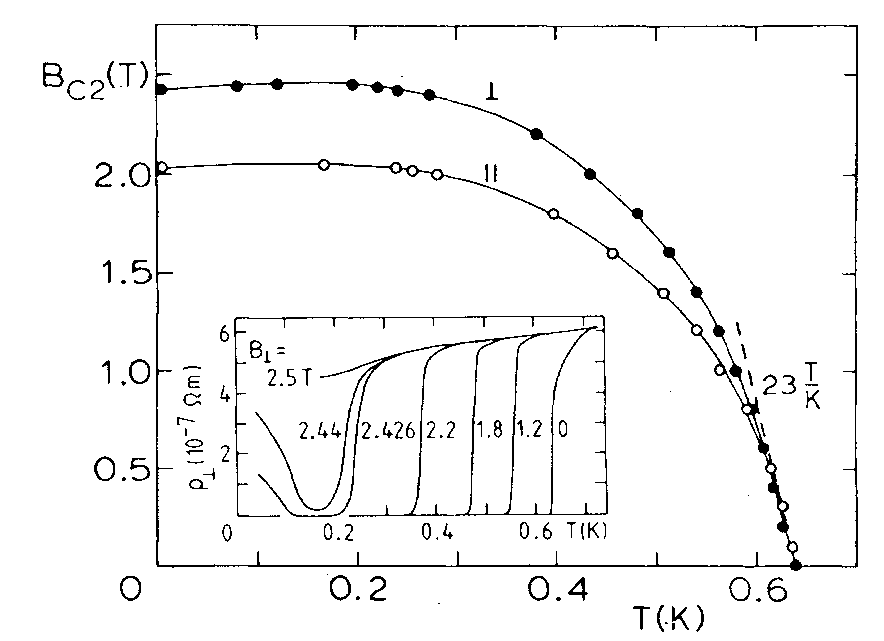}
	\end{center}
	\caption{Upper critical magnetic field $B_{\rm c2}$ vs $T$ of a CeCu$_2$Si$_2$ single crystal for fields applied within ($\parallel$) and perpendicular to ($\perp$)  the Ce planes as obtained from $\rho(T)$, measured parallel to the respective field. Only a moderate anisotropy, but a giant initial slope at $T_{\rm c}$ is found for  $B_{\rm c2}$(T). Note the shallow maximum of $B_{\rm c2}$(T) near $T=0.15$~K as reflected in the inset by the reentrant $\rho(T)$ behavior for $B\geq2.4$~T. Reproduced with permission from \onlinecite{Assmus1984}.}
	\label{Fig2}
\end{figure}	

The jump height at the superconducting transition, $\Delta C/T_{\rm c}$, is comparable to the Sommerfeld coefficient, extrapolated to $T=0$, $\gamma_0 = C(T\rightarrow0)/T \approx1$~J/mol~K$^2$ (Fig.~\ref{Fig1}(b)). This not only proved bulk superconductivity, but also led to the conclusion that the Cooper pairs are formed by heavy-mass quasiparticles \cite{Steglich1979} and to the term `heavy-fermion superconductivity' \cite{Rauchschwalbe1982}. In fact, if the superconductivity were solely carried by the coexisting light conduction electrons, the jump in the electronic specific-heat coefficient at $T_{\rm c}$ would have been so tiny that within the scatter of the data it would not be resolvable in Fig.~\ref{Fig1}(b). It should be noted that recent theoretical considerations have shown that in order to form Cooper pairs in CeCu$_2$Si$_2$, a very large kinetic energy cost, exceeding the binding energy by a factor as high as 20,  is necessary to overcompensate the  similarly large exchange energy between the paired  heavy quasiparticles \cite{Stockert2011}. The large kinetic-energy cost has been interpreted in terms of a transfer of single-electron spectral weight to energies above a Kondo-destruction energy scale at the QCP, $T^*$, that is nonzero but small compared to the bare Kondo scale \cite{Stockert2011}, see Sec.~\ref{QCPSec}.

The two polycrystalline samples exploited in Fig.~\ref{Fig1}(b) were prepared and annealed in the same way. Nevertheless, their specific-heat values were found to be significantly different. These variations of physical properties from one sample to the other added to the severe skepticism \cite{Hull1981,Schneider1983} which existed throughout the first few years after the report of bulk superconductivity in CeCu$_2$Si$_2$ \cite{Steglich1979}, subsequently confirmed by other groups \cite{Ishikawa1983,Aliev1983a,Aliev1984,Onuki1984}. The cause for these ‘sample dependences’, see also \cite{Stewart1983,Aliev1983a}, was resolved only many years later by a thorough study of the chemical phase diagram \cite{MullerThesis,Steglich2001}, and the observation of a quantum critical point (QCP) that is located inside the narrow homogeneity range \cite{Lengyel2011,Steglich2001}, see Sec.~\ref{QCPSec}. The above-mentioned skepticism could be overcome after a few years, when high-quality single crystals of CeCu$_2$Si$_2$ were prepared \cite{Assmus1984,Onuki1984} and found to show even more pronounced superconducting phase transition anomalies compared to polycrystals. The upper critical field curve $B_{\rm c2}(T)$ of such a single crystal is displayed in Fig.~\ref{Fig2}. It reveals:\\ 
(i) Only a small anisotropy between the field being applied parallel and perpendicular to the basal tetragonal plane [inset of Fig.~\ref{Fig1new}], contrasting a pronounced anisotropy in the electrical resistivity \cite{Schneider1983}.\\
 (ii) A shallow maximum around $T=0.15$~K (inset) which seems to correspond to a low-temperature hump in $C(T)/T$ \cite{Bredl1984}, also observed for CeAl$_3$ \cite{Flouquet1982,Bredl1984,Steglich1985}, which was ascribed to the opening of a partial coherence gap in the  $4f$-quasiparticle density of states at the Fermi level, see Table~\ref{table:table1}. Later this hump was ascribed as being related to antiferromagnetic  correlations \cite{Steglich1996,Stockert2004}. For UBe$_{13}$ too, a broad peak in $C(T)/T$ at $T_{\rm L}\approx0.6$~K had been detected \cite{Rauchschwalbe1987a,Rauchschwalbe1987b}  and subsequently identified \cite{Kromer1998,Kromer2000} as the precursor of an anomaly indicating a continuous phase transition at $T_{\rm c2}$ below the superconducting $T_{\rm c1}$ discovered for           
(U$_{1-x}$Th$_x$)Be$_{13}$ in the critical concentration range $0.019\leq x \leq 0.045$ \cite{Ott1985}. The nature of this lower-lying phase transition has yet to be resolved \cite{Steglich2016}. While ultrasound-attenuation results \cite{Batlogg1985} hint at a SDW transition, pressure studies \cite{Lambert1986} and results of the lower critical field \cite{Rauchschwalbe1987a} highlight a superconducting nature of the transition at $T_{\rm c2}$.\\
  (iii) A giant initial slope at $T_{\rm c}$ which supports the massive nature of the Cooper pairs as inferred from the giant jump anomaly $\Delta C/T_{\rm c}$.\\
  (iv) Quite a strong Pauli limiting effect in the low-temperature regime for both field configurations. This discards odd-parity (spin-triplet) pairing as observed in superfluid $^3$He \cite{Leggett1975} and originally assumed for heavy-fermion superconductors \cite{Anderson1984}. A spatially modulated superconducting state in CeCu$_2$Si$_2$ at very low temperature close to the upper critical field was recently proposed based on Cu-NMR results \cite{Kitagawa2018}.

A dc Josephson effect with a critical pair current of ordinary size was observed on a weak link between polycrystalline CeCu$_2$Si$_2$ and Al \cite{Steglich1985}. This as well as Knight shift results from $^{29}$Si NMR \cite{Ueda1987} lent further support to even-parity (spin-singlet) pairing in CeCu$_2$Si$_2$. 

At around the same time, theorists proposed $d$-wave superconductivity mediated by antiferromagnetic  spin fluctuations \cite{Miyake1986,Scalapino1986}. 
These theoretical studies extend the theory of ferromagnetic paramagnons developed in the $^3$He context to the antiferromagnetic case, 
but the Kondo effect responsible for the heavy mass was not addressed.
In more recent years, the Kondo effect has been incorporated into the study of heavy-fermion quantum criticality \cite{Gegenwart2008}, with an emphasis on the notion of Kondo destruction \cite{Si01.1,Col01.1}.
A corresponding theory for quantum-criticality-driven superconductivity in Kondo-lattice models has been advanced \cite{Hu2021}.

The discovery of heavy-fermion superconductivity in the cubic compound UBe$_{13}$ \cite{Ott1983} proved this phenomenon to be general and not restricted to a single material. Thereafter, UPt$_3$ \cite{Stewart1984b}, URu$_2$Si$_2$ \cite{Schlabitz1984,Schlabitz1986,Palstra1985,Maple1986}, U$_2$PtC$_2$ \cite{Meisner1984}, UNi$_2$Al$_3$ \cite{Geibel1991Ni}, and UPd$_2$Al$_3$ \cite{Geibel1991Pd} were found to be heavy-fermion superconductors, too. They were followed by the pressure-induced Ce-based heavy-fermion superconductors CeCu$_2$Ge$_2$ \cite{Jaccard1992}, CeRh$_2$Si$_2$ \cite{Movshovich1996}, CeIn$_3$ and CePd$_2$Si$_2$ \cite{Mathur1998}. In the years after 2000, many of the Ce-based tetragonal, so-called 115 materials, which are obtained by increasing the $c/a$ ratio of cubic CeIn$_3$ by inserting an additional layer of $T$In$_2$ ($T$: Co, Rh or Ir), as well as the related 218 and 127 compounds were shown to be heavy-fermion superconductors \cite{Sarrao2007,Thompson2012}. One of the Pu-based isostructural compounds, PuCoGa$_5$, exhibits the record $T_{\rm c}=18.5$~K for this class of superconductors \cite{Sarrao2002}. Its Rh homologue PuRhGa$_5$ \cite{Wastin2003} as well as NpPd$_5$Al$_2$ \cite{Aoki2009} also show enhanced $T_{\rm c}$ values of 8.7~K and 4.9~K, respectively. The discovery of heavy-fermion superconductivity in the noncentrosymmetric compound CePt$_3$Si \cite{Bauer2004} stimulated the search for noncentrosymmetric heavy-fermion as well as weakly-correlated superconductors \cite{Bauer2012,Smidman2017}, and resulted in several Ce-based counterparts. Such a lack of inversion symmetry allows for a mixing between even- and odd-parity pairing states \cite{Gorkov2001}. In the case of CeRh$_2$As$_2$ which has a locally noncentrosymmmetric crystal structure, two-phase superconductivity has been reported very recently, along with a proposal for a field-induced transition between an even parity phase at low fields, and an odd parity phase at elevated fields \cite{Khim2021}. Two different superconducting  phases in the presence of weak antiferromagnetic order have already been established earlier for UPt$_3$ \cite{Joynt2002}, and multifaceted behavior has been reported for thoriated UBe$_{13}$ \cite{Ott1985,Heffner1990,Oeschler2003} as well as URu$_2$Si$_2$, exhibiting a hidden-order phase \cite{Mydosh2020}. All of the three latter materials show a superconducting state with broken time-reversal symmetry \cite{Luke1993,Schemm2014,Heffner1990,Schemm2015}.  

There are only two Yb-based heavy-fermion superconductors known so far. $\beta$-YbAlB$_4$ with $T_{\rm c}=80$~mK \cite{Nakatsuji2008} is an intermediate-valence compound showing quantum criticality without tuning \cite{Matsumoto2011}. YbRh$_2$Si$_2$ \cite{Schuberth2016,Nguyen2021,Schuberth2022,Shan2022} exhibits an antiferromagnetic QCP at $B\approx0$ which is induced by nuclear spin order (below $T_{\rm A} = 2.3$~mK). The latter strongly competes with the primary $4f$-electronic order ($T_{\rm N}=70$ mK) and causes the emergence of heavy-fermion superconductivity at ultra-low temperatures, $T_{\rm c}=2$~mK. As shown by \cite{Schuberth2022}, measurements of the Meissner effect point to the existence of bulk superconductivity up to magnetic fields of the order of $B=40$~mT (about two-thirds of $B_N$, the critical field designating the Kondo-destruction QCP \cite{Custers2003}). Furthermore recent resistivity investigations suggest that at such elevated fields superconductivity may be of the spin-triplet variety \cite{Nguyen2021}, which is theoretically supported based on unconventional superconductivity driven by Kondo destruction at magnetic-field-induced quantum criticality in the presence of an effective Ising spin anisotropy  \cite{Hu2021.2}. Correlated Pr-based superconductors were also found. PrOs$_4$Sb$_{12}$ shows a heavy-fermion normal-state and superconducting properties due to  dominant quadrupolar rather than dipolar fluctuations \cite{Maple2002,Rotundu2004}, while PrTi$_2$Al$_{20}$, PrV$_2$Al$_{20}$ and PrIr$_2$Zn$_{20}$ are quadrupolar Kondo-lattice systems exhibiting superconductivity and quadrupolar order \cite{Sakai2012,Tsujimoto2014,Onimaru2011}. 

A few heavy fermion superconductors are prime candidates for odd-parity pairing, i.e., the ferromagnetic compounds UGe$_2$ \cite{Saxena2000}, URhGe \cite{Levy2005} and UCoGe \cite{Huy2007,Hattori2012} as well as UPt$_3$ \cite{Tou1998} and UNi$_2$Al$_3$ \cite{Ishida2002}. Also being discussed is UTe$_2$ \cite{Ran2019,Aoki2019}. It has been suggested to be a chiral topological superconductor \cite{Jiao2020}, for which the role of Kondo and RKKY interactions in the magnetic correlations and superconductivity has been discussed \cite{Thomas2020,Duan2020,Knafo2021,Duan2021,Chen2021}. 

In concluding this survey, 
we can state that currently about fifty heavy-fermion superconductors are known. Most of these materials are discussed in \cite{Pfleiderer2009}. They are complemented by the already mentioned compounds $\beta$-YbAlB$_4$, Pr(Ti,V)$_2$Al$_{20}$, PrIr$_2$Zn$_{20}$ YbRh$_2$Si$_2$, UTe$_2$, and CeRh$_2$As$_2$. The majority of heavy-fermion superconductors are  believed to have anisotropic even-parity  Cooper pairing. In the following section, we present the early evidence for single-band $d$-wave superconductivity in CeCu$_{2}$Si$_{2}$ down to about $T=0.1$ K, see also \cite{Stockert2012}.

\section{Evidence for $d$-wave superconductivity in C\lowercase{e}C\lowercase{u}$_2$S\lowercase{i}$_2$ above 0.1~K}
\subsection{Phase diagram}

\begin{figure}[t]
	\begin{center}
		\includegraphics[width=0.85\columnwidth]{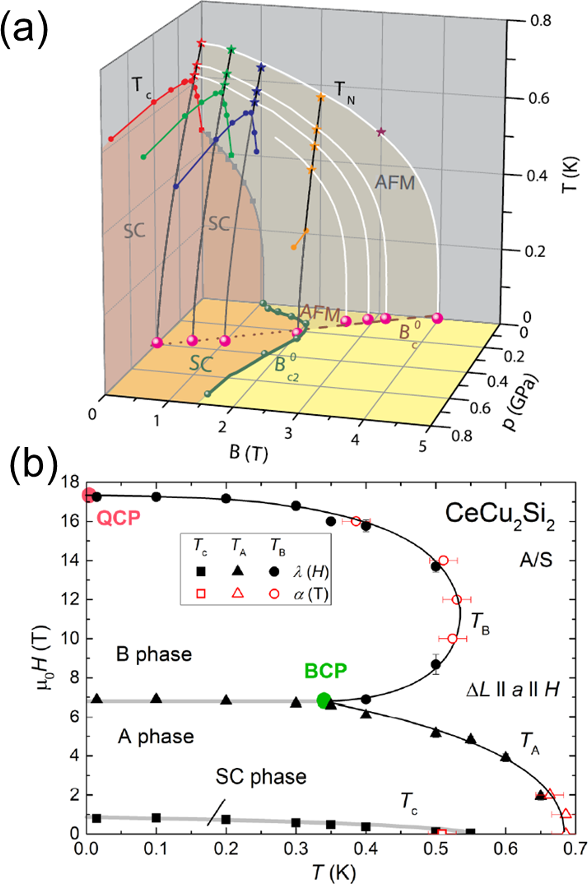}
	\end{center}
	\caption{(a) Temperature-pressure-magnetic field phase diagram of a single crystal of $A/S$-type CeCu$_2$Si$_2$. Reproduced with permission from \onlinecite{Lengyel2011}. (b) Magnetic field - temperature diagram of single crystalline CeCu$_2$Si$_2$, where positions of the field-induced bicritical point (BCP) and QCP are also displayed. Reproduced with permission from \onlinecite{Weickert2018}. }
	\label{Fig3}
\end{figure}	

\begin{figure}[t]
	\begin{center}
		\includegraphics[width=0.9\columnwidth]{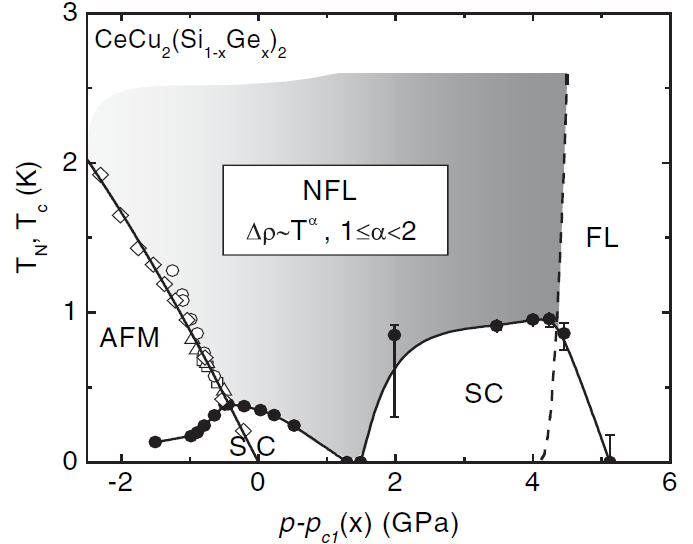}
	\end{center}
	\caption{Temperature - pressure phase diagram of CeCu$_2$(Si$_{1-x}$Ge$_2$)$_2$, which exhibits two superconducting domes (for $x=0.1$), one centered around a lower pressure $p_{\rm c1}$ associated with an antiferromagnetic QCP, while the dome at higher pressures is near a possible valence transition. The diamonds, circles, triangles  and squares correspond to compositions with $x=0.25, 0.1, 0.05,$ and $0.01$, respectively. The dashed line displays the anticipated line of first-order valence transitions, ending  in a critical point somewhere between 10 and 20 K. Solid lines are guides to the eye. Reproduced with permission from \onlinecite{Yuan2006}.}
	\label{twodome}
\end{figure}

One of the major distinguishing features which sets CeCu$_2$Si$_2$ apart from previously-known BCS superconductors is the close proximity between magnetism and superconductivity in the phase diagram, where both are due to the same localized $4f$-electrons. This is reflected in the observation that slight tuning of the Cu:Si ratio within the homogeneity range can lead to crystals with ground states which are entirely antiferromagnetic ($A$-type), superconducting ($S$-type) or exhibit both superconductivity and magnetism ($A/S$-type) \cite{Steglich1996,Seiro2010}. While within the context of BCS theory, magnetism and superconductivity are generally considered antagonistic, superconductivity on the border of magnetism is a common feature of broad classes of unconventional superconductors \cite{Norman2011,Norman2013,Stewart2017} including heavy-fermion superconductors \cite{Pfleiderer2009,Steglich2016}, cuprates \cite{Lee2006,Proust2019}, iron-based pnictides and chalcogenides \cite{Stewart2011,Si2016}, organic superconductors \cite{Maple2004,Lang2003,Kanoda2008} and twisted graphene superlattices \cite{Cao2018,Andrei2020}, and may be related to the occurrence of  Cooper pairs with a magnetically driven pairing interaction \cite{Scalapino2012}, rather than the conventional electron-phonon pairing mechanism.

The temperature-pressure-magnetic field diagram of an $A/S$-type single crystal is displayed in Fig.~\ref{Fig3}(a) \cite{Lengyel2011}. At ambient pressure, two zero-field phase transitions can be detected in specific heat measurements corresponding to an antiferromagnetic transition at $T_{\rm N}=0.69$~K and a subsequent superconducting transition at $T_{\rm c}=0.46$~K. The application of moderate pressure rapidly suppresses $T_{\rm N}$, while $T_{\rm c}$ shows a slight increase, and once $T_{\rm N}$ is suppressed below $T_{\rm c}$, no antiferromagnetic transition is observed. When a magnetic field is applied, both $T_{\rm N}$ and $T_{\rm c}$ are suppressed, but the more rapid decrease of $T_{\rm c}$ with field allows for $T_{\rm N}$  (extrapolated to $B=0$) to be tracked as a function of pressure to lower temperatures. From extrapolating the positions of $B^0_{c}$ (where  $T_{\rm N}$ vanishes)  for fixed values of pressure,  a line of QCPs is inferred to lie in the  zero-temperature  pressure-field phase diagram shown in  Fig.~\ref{Fig3}(a).  $B^0_{c}(p)=0$ at  $p_c=0.39$~GPa, which is almost twice as large as the pressure where  $B^0_{c2}$ vs $p$  exhibits a local maximum (see Sec.~\ref{PersSec}). $p_c$ can be forced to vanish, if the composition of the (homogeneous) sample becomes slightly more enriched  by Cu (i.e., by reducing the average unit-cell volume). Although the ambient-pressure, zero-field QCP is masked by superconductivty, its nature can be well explored by studying the low-temperature normal state of such an $S$-type sample induced by applying a small external magnetic field, see Sec.~\ref{QCPSec}. 

Detailed measurements of the elastic constants, thermal expansion, and magnetostriction revealed the presence of a field-induced `B' phase, in addition to the magnetic `A' phase found at low fields \cite{Bruls1994,Weickert2018}. The field-temperature phase diagram is displayed in Fig.~\ref{Fig3}(b), where there are second-order lines between the paramagnetic state and both the `A' and `B' phases, while going from the `A' to `B' phase corresponds to a first-order transition, leading to a bicritical point in the phase diagram between these phases \cite{Weickert2018}. Measurements to very low temperatures and high fields show the suppression of the `B' phase to zero temperature in applied fields of around 17~T, giving rise to a field-induced QCP. The nature of the transition from `A' to `B' phase is still to be determined, where the small change in magnetization between the two phases suggests that `B' (like `A', see below) also corresponds to a spin-density-wave (SDW) phase \cite{Tayama2003,Weickert2018}.

The shape of the superconducting region in the temperature-pressure phase diagram of $S$-type CeCu$_2$Si$_2$ is quite unusual compared to other heavy-fermion superconductors \cite{Mathur1998,Knebel2006}, namely at low and moderate pressures $T_{\rm c}$ does not change rapidly with pressure, while at higher pressures it reaches a maximum at around 4~GPa, well away from the point where magnetism is suppressed \cite{Bellarbi1984,Thomas1993,Yuan2003,Yuan2006}. Remarkably, upon substituting 10 at\% of Si by  Ge, which substantially reduces  $T_{\rm c}$,  it is found that this actually results in two superconducting domes in the phase diagram, as shown in  Fig.~\ref{twodome}, where one is centered around the antiferromagnetic QCP, and another with a higher maximum $T_{\rm c}$ occuring at higher pressures \cite{Yuan2003,Yuan2006}. It has been suggested that these two domes correspond to superconductivity with different unconventional pairing mechanisms, with the low pressure dome corresponding to magnetically driven superconductivity and the high pressure dome driven by charge (valence) fluctuations \cite{Yuan2003,Holmes2004}. A similar phase diagram with two superconducting domes was reported for the (Pu,Co)-based 115 systems by \cite{Bauer2012}. Here, the higher $T_{\rm c}$ of PuCoGa$_5$ (18.5~K) compared to PuCoIn$_5$  (2.5~K) was ascribed to the superconductivity of the former arising from a valence instability, while that of the latter was associated with a magnetic quantum critical point.

\begin{table*}[tb]
\caption{Chronology of discoveries and early studies on heavy fermions, heavy-fermion superconductivity, and related topics (1969-1989).The following abbreviations are used: PM: paramagnetic, AF(O): antiferromagnetic (order), SDW: spin-density wave, CDW: charge-density wave, MF: mean field, FL: Fermi liquid, HF: heavy fermion, KE: Kondo effect, RLM: resonance level model \cite{Schotte1975}, I: interpretation.}
\label{table:table1}
\begin{ruledtabular}
 \begin{tabular}{c c c c }
\textbf{Year} &\textbf{Discovery/Achievement}  &\textbf{Material}  &\textbf{Reference}   \\
\hline\hline\\[-2ex]

1969  &First synthesis   &CeCu$_2$Si$_2$   & \onlinecite{Rieger1969} \\ 
\hline & & & \\[-1.0em]
1969 &Superconductivity, $T_c$ = 1.47 K  &U$_2$PtC$_2$   & \onlinecite{Matthias1969}   \\ 
\hline & & & \\[-1.0em]
1971 & Fe/Cr-derived specific heat  & \textbf{Cu}(Fe,Cr)   & \onlinecite{Triplett1971}   \\
 & $\Delta C(T)=\gamma T$ at $T\ll T_{\rm K}$, $\gamma\approx1(16)$~J/mol~K$^2$  & 80 (20-50) ppm   &    \\ 
 \hline & & & \\[-1.0em]
 1972 & Superfluidity & Liquid $^3$He & \onlinecite{Osheroff1972a,Osheroff1972b}\\ 
 \hline & & & \\[-1.0em]
 1974 & Theory of local FL of $S=\frac{1}{2}$ Kondo ion & & \onlinecite{Nozieres1974} \\ 
 \hline & & & \\[-1.0em]
 1975 & Superconducting transition at $T_c$ = 0.97~K  & UBe$_{13}$ & \onlinecite{Bucher1975}\\
 & $T_c$ decreases by  $30\%$ in $B=6$~T. \textbf{I:} due to U - filaments &&\\ 
 \hline & & & \\[-1.0em]
 1975& Heavy FL; $\gamma=1.62$~J/mol~K$^2$ & CeAl$_3$ & \onlinecite{Andres1975} \\
 & \textbf{I:} due to $4f$-virtual bound state & & \\ 
  \hline & & & \\[-1.0em]
 1975 & Treatment of KE by renormalization group & & \onlinecite{Wilson1975} \\ 
 \hline & & & \\[-1.0em]
 1975 & Theory of superfluid phases & Liquid $^3$He & \onlinecite{Leggett1975} \\ 
 \hline & & & \\[-1.0em]
 1976 & Magnetic properties & CeCu$_2$Si$_2$ & \onlinecite{Sales1976} \\
  & \textbf{I:} Intermediate-valence compound & & \\ 
  \hline & & & \\[-1.0em]
  1978 & $T_{\rm K}=5$~K, $T_{\rm N}=3.9$~K, $\gamma_{\rm AF}=0.135$~J/mol~K$^2$  &CeAl$_2$ & \onlinecite{Bredl1978}\\
& KE/AFO treated by RLM/MF: $\gamma_{\rm PM}=1.7$~J/mol~K$^2$ & & \\ 
\hline & & & \\[-1.0em]
 1978 & Superconducting transition at $T_{\rm c}\approx0.5$~K in resistivity  & CeCu$_2$Si$_2$ & \onlinecite{Franz1978} \\ 
  & and susceptibility  \textbf{I:} due to spurious phase(s) & &\\ 
 \hline & & & \\[-1.0em]
1979 & Bulk superconductivity, $T_{\rm c}\approx 0.6$~K (first HF superconductor) & CeCu$_2$Si$_2$ & \onlinecite{Steglich1979} \\
 & $\gamma\approx1$~J/mol~K$^2$, heavy fermions  (introduction of the term ``HF'') & &\\ 
 \hline & & & \\[-1.0em]
 1982 & Lower and upper critical fields & CeCu$_2$Si$_2$ & \onlinecite{Rauchschwalbe1982} \\
 & Meissner effect, strong Pauli limiting, \textbf{I:} even-parity pairing & & \\ 
 \hline & & & \\[-1.0em]
 1983 & HF superconductivity ($T_{\rm c}\approx 0.85$~K, $\gamma\approx1.1$~J/mol~K$^2$) & UBe$_{13}$ & \onlinecite{Ott1983} \\ 
 \hline & & & \\[-1.0em]
 1983 & Suppression of superconductivity by $\approx1\%$ impurity substitution  & CeCu$_2$Si$_2$ & \onlinecite{Spille1983} \\
 & \textbf{I:} Unconventional superconductivity & & \\ 
 \hline & & & \\[-1.0em]
  1984 & HF superconductivity in single crystals & CeCu$_2$Si$_2$ & \onlinecite{Assmus1984} \\ 
  &  &  & \onlinecite{Onuki1984} \\ 
 \hline & & & \\[-1.0em]
   1984 & HF superconductivity ($T_c=0.5$~K, $\gamma=0.4$~J/mol~K$^2$) & UPt$_3$ & \onlinecite{Stewart1984b} \\ 
  &  &  & \onlinecite{Onuki1984} \\ 
 \hline & & & \\[-1.0em]
 1984 & Hump in $C(T)/T$, \textbf{I:} due to Kondo lattice coherence & CeCu$_2$Si$_2$/CeAl$_3$ & \onlinecite{Bredl1984}\\ 
 \hline & & & \\[-1.0em]
 1984 & $C(T)\sim T^3$($T\ll T_{\rm c}$)& UBe$_{13}$ & \onlinecite{Ott1984} \\ 
   &\textbf{I:} gap point nodes, $p$-wave superconductivity &  &  \\ 
 \hline & & & \\[-1.0em]
 1984 & NMR: $1/T_1\sim T^3$, \textbf{I:} gap line nodes & (U$_{1-x}$Th$_x$)Be$_{13}$  & \onlinecite{MacLaughlin1984} \\ 
 \hline & & & \\[-1.0em]
 1984 & Theory of superconductivity in Kondo lattice  & & \onlinecite{Razafimandimby1984}\\
 & by Gr\"{u}neisen-parameter coupling & & \\
 \hline & & & \\[-1.0em]
 1984 & Theory of triplet pairing in HF superconductors & & \onlinecite{Anderson1984}\\
  \hline & & & \\[-1.0em]
1984 & HF superconductivity ($T_{\rm c}\approx 1.5$~K, $\gamma\approx0.075$~J/mol~K$^2$)& U$_2$PtC$_2$ & \onlinecite{Meisner1984} \\
    \hline & & & \\[-1.0em]
1984 & HF superconductivity ($T_{\rm c}\approx 0.8-1.5$~K, $\gamma\approx0.07$~J/mol~K$^2$)& URu$_2$Si$_2$ & \onlinecite{Schlabitz1984,Schlabitz1986} \\
& MF-type transition at $T_0=17.5$~K, \textbf{I:} into SDW/CDW & & \onlinecite{Palstra1985} \\
 &   &   & \onlinecite{Maple1986} \\
      \hline & & & \\[-1.0em] 
 1985   & dc-Josephson effect across  CeCu$_2$Si$_2$/Al weak link:   & CeCu$_2$Si$_2$   & \onlinecite{Steglich1985} \\
    &  ordinary critical pair current size   &   &  \\
    \hline & & & \\[-1.0em]  
1985 & Second transition below $T_{\rm c}$, \textbf{I:} Unconventional superconductivity & (U$_{1-x}$Th$_x$)Be$_{13}$ & \onlinecite{Ott1985} \\
     \hline & & & \\[-1.0em]  
 1985 & Second transition below $T_{\rm c}$, \textbf{I:} SDW transition & (U$_{1-x}$Th$_x$)Be$_{13}$ & \onlinecite{Batlogg1985} \\
     \hline & & & \\[-1.0em]   
 1986 & Evidence for two superconducting states   & (U$_{1-x}$Th$_x$)Be$_{13}$ & \onlinecite{Lambert1986} \\
     \hline & & & \\[-1.0em]     
1986 & Penetration depth: $\lambda(T)\sim T^2$($T\ll T_{\rm c}$), \textbf{I:} gap point nodes &  UBe$_{13}$ & \onlinecite{Gross1986} \\
\hline & & & \\[-1.0em]
1986 & Theory of even-parity pairing caused by spin fluctuations & & \onlinecite{Miyake1986} \\
\hline & & & \\[-1.0em]
1986 & Theory of $d$-wave pairing near an SDW instability & &\onlinecite{Scalapino1986} \\
     \hline & & & \\[-1.0em]  
 1987 & Evidence for two coexisting superconducting order parameters & (U$_{1-x}$Th$_x$)Be$_{13}$ & \onlinecite{Rauchschwalbe1987b} \\
\hline & & & \\[-1.0em]
1988 & dHvA oscillations: direct observation of HFs & UPt$_3$ & \onlinecite{Taillefer1988} \\
\hline & & & \\[-1.0em]
1988 & Penetration depth: $\lambda(T)\sim T^2$($T\ll T_{\rm c}$), \textbf{I:} gap  nodes &  UPt$_{3}$, CeCu$_2$Si$_2$  & \onlinecite{Gross1998} \\
\hline & & & \\[-1.0em]
1989 & Second transition below $T_{\rm c}$, \textbf{I:} Unconventional superconductivity & UPt$_3$ & \onlinecite{Fisher1989} \\
\hline & & & \\[-1.0em]
1989 & Weak AFO, decrease of magnetic Bragg intensity below $T_{\rm c}$  & UPt$_3$ & \onlinecite{Aeppli1989} \\
\hline & & & \\[-1.0em]
1989 & Theory on broken symmetry in an unconventional superconductor &  & \onlinecite{Hess1989} \\
& model for double transition in UPt$_3$ & & \\
\hline & & & \\[-1.0em]
1989 & Phenomenological theory of multiple pairing states & UPt$_3$  & \onlinecite{Machida1989} \\
\end{tabular}
\end{ruledtabular}
\end{table*}

\begin{table*}[tb]
\centering
\caption{A summary of experimental probes of the superconducting gap structure of CeCu$_2$Si$_2$, together with proposed theories for the superconducting pairing state.}
\label{table:table2}
\begin{ruledtabular}
 \begin{tabular}{c c c c }
\multicolumn{4}{c}{\large{\textbf{Experiments}}} \\
\hline\hline\\
\textbf{Probe}   &\textbf{Results}  &\textbf{Interpretation} &\textbf{Reference}   \\
\hline \\[2ex]
Resistivity under field & Paramagnetic limiting of $B_{\rm c2}$ & Singlet pairing & \onlinecite{Assmus1984} \\ \\

Specific heat  & $C\sim T^3$& -- & \onlinecite{Steglich1985a} \\ \\

NMR Knight shift  & Knight shift decrease below $T_{\rm c}$ & Singlet pairing & \onlinecite{Ueda1987} \\ \\

Penetration depth  & $\lambda\sim T^2$ & Gap nodes & \onlinecite{Gross1998} \\ \\

Point contact spectroscopy  & Flat $dV/dI$ & Nodeless gap & \onlinecite{Wilde1994} \\ \\

Cu-NQR   & $1/T_1\sim T^3$ & Gap line nodes & \onlinecite{Ishida1999} \\ 
  &  &  & \onlinecite{Fujiwara2008} \\ \\
Inelastic neutron scattering   & Peak in magnetic response  & Sign-changing  & \onlinecite{Stockert2011} \\ 
&below $T_{\rm c}$&order parameter&\\ \\

Field angle dependent resistivity   & 4-fold $B_{\rm c2}(\phi)$ & $d_{xy}$ state & \onlinecite{Vieyra2011} \\ \\

Specific heat ($T<0.1$~K)   & Exponential $C(T)$ as $T\rightarrow0$  & Two nodeless gaps & \onlinecite{Kittaka2014,Kittaka2016} \\ \\

Scanning tunneling microscopy   & Spectra analysis  & Nodal + nodeless gaps & \onlinecite{Enayat2016} \\ \\

Penetration depth  ($T<0.1$~K)   & Exponential $\lambda(T)$ as $T\rightarrow0$  & Nodeless gap & \onlinecite{Yamashita2017}  \\ 
&&& \onlinecite{Takenaka2017} \\ 
&&& \onlinecite{Pang2018} \\  \\

Thermal conductivity ($T<0.1$~K)   & Vanishing $\kappa/T$ as $T\rightarrow0$  & Nodeless gap & \onlinecite{Yamashita2017} \\ \\

Cu-NQR ($T<0.1$~K)   & Exponential $1/T_1$ as $T\rightarrow0$  & Nodeless gap & \onlinecite{Kitagawa2017} \\ \\

Small angle neutron scattering   & Form factor analysis  & Two nodeless gaps & \onlinecite{Campillo2021} \\ \\ \\

\end{tabular}
 \begin{tabular}{c c c c}
\multicolumn{4}{c}{\large{\textbf{Theory}}} \\ 
\hline\hline\\
\textbf{Theory}   &\textbf{Gap structure} &\textbf{Sign-change?}  &\textbf{Reference}   \\
\hline \\[2ex]
Loop nodal $s_{+-}$ state & Nodal & $\checkmark$(interband) & \onlinecite{Ikeda2015} \\ \\
$d+d$ pairing & Nodeless & $\checkmark$(intraband) & \onlinecite{Nica2017} \onlinecite{Nica2021} \\ \\
Multipole mediated $s$-wave & Nodeless & $\times$ & \onlinecite{Tazai2018,Tazai2019} \onlinecite{Tazai2019} \\\\
$s_{+-}$ state & Nodal, nodeless & $\checkmark$(interband) & \onlinecite{Li2018} \\ \\
\end{tabular}
\end{ruledtabular}
\end{table*}

\subsection{Origin of the A-phase in CeCu$_2$Si$_2$}
\label{AOrigin}

Although the relative increase of the electrical resistivity below the ordering temperature $T_{\rm N}$ suggested the opening of an excitation gap in CeCu$_2$Si$_2$ due to a SDW-type of magnetic order \cite{Gegenwart1998}, direct evidence for such a scenario was lacking for a long time. 
The first  indications for antiferromagnetic order as the characteristic of the A-phase came from NMR \cite{Nakamura1988} and muon-spin relaxation ($\mu$SR) measurements \cite{Uemura1988,Uemura1989} in the late 1980s, both detecting a static magnetic field (at the muon site or the nuclear site, respectively) in the ordered state. In these measurements even an incommensurate type of magnetic order in CeCu$_2$Si$_2$ was proposed because of the distribution of local magnetic fields detected. Interestingly, while pronounced phase transition anomalies at $T_{\rm N}$ were found in both elastic-constant and thermal-expansion measurements \cite{Bruls1994}, no corresponding feature was seen in the magnetic susceptibility for a long time, until a cusp-like anomaly could eventually be resolved in the susceptibility when monitored with the aid of a high-resolution Faraday magnetometer \cite{Tayama2003}. 

\begin{figure}[t]
	\begin{center}
		\includegraphics[width=0.75\columnwidth]{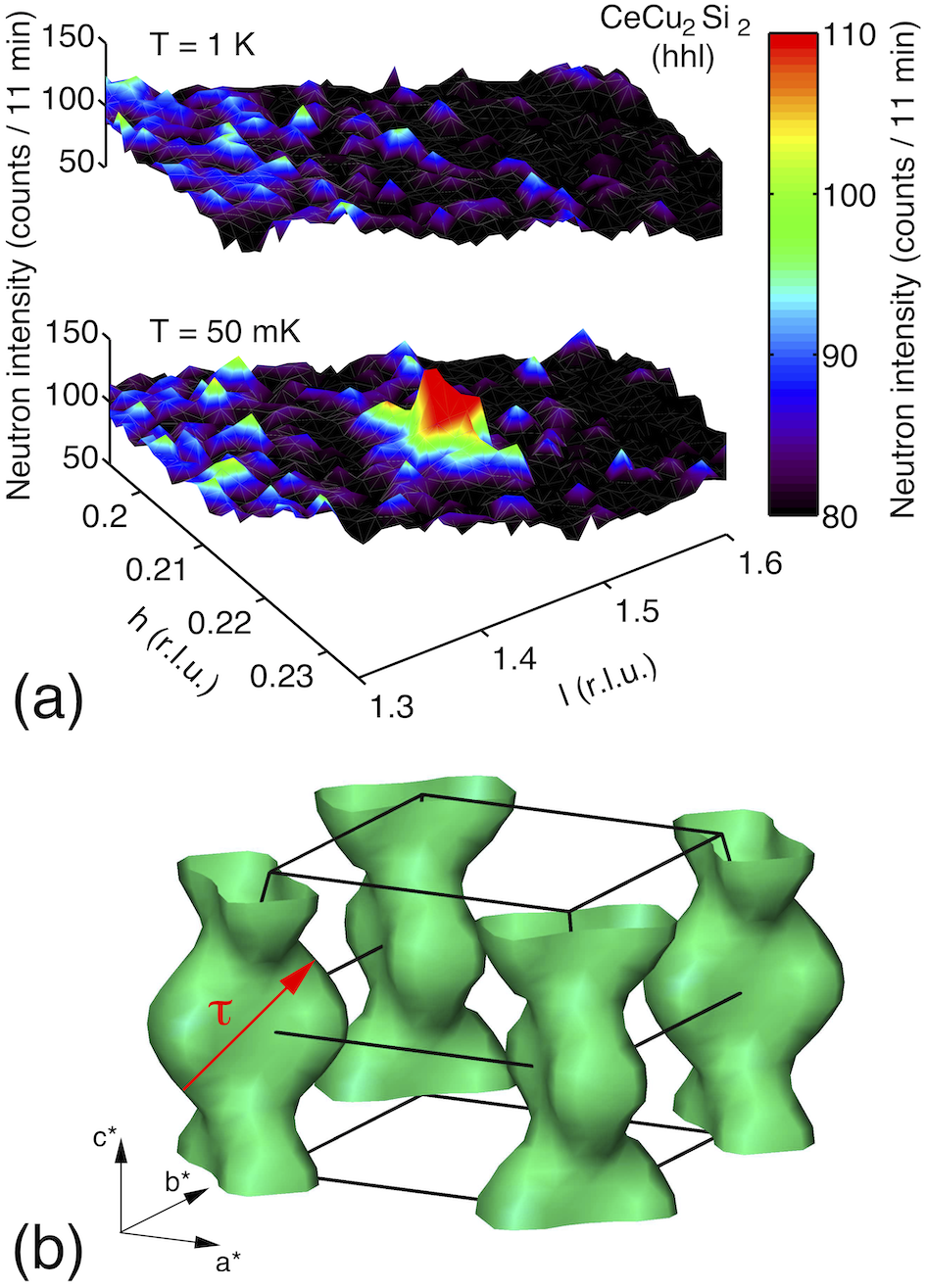}
	\end{center}
	\caption{(a) Neutron-diffraction intensity map of the reciprocal $(h~h~l)$ plane around $\textbf{Q} = (0.21~0.21~1.45)$ in A-phase CeCu$_2$Si$_2$ at $T = 50$\,mK and $1$\,K. (b) Main heavy Fermi surface sheet in CeCu$_2$Si$_2$ indicating the columnar nesting with wave vector $\boldsymbol\tau$. Reproduced with permission from \onlinecite{Stockert2004}.}
	\label{Fig3_1}
\end{figure}

In 1997, antiferromagnetic order was observed in the reference compound CeCu$_2$Ge$_2$ using single crystal neutron diffraction \cite{Krimmel1997b} which later could be related to the nesting properties of the Fermi surface \cite{Zwicknagl2007}. In order to unravel the nature of the A-phase in pure CeCu$_2$Si$_2$, an approach to study the magnetic order in the Ge-substituted system CeCu$_2$(Si$_{1-x}$Ge$_x$)$_2$ was chosen. Starting from pure CeCu$_2$Ge$_2$ the antiferromagnetic order was followed in CeCu$_2$(Si$_{1-x}$Ge$_x$)$_2$ with decreasing Ge content. Initially the incommensurate order in CeCu$_2$(Si$_{1-x}$Ge$_x$)$_2$ was detected only for $x \ge 0.6$ in neutron powder diffraction \cite{Knebel1996,Krimmel1997a}. However, measurements in powder samples with lower Ge concentrations were unsuccessful since the ordering temperature as well as the magnetically ordered moment are largely reduced for samples with low Ge content. Until the early 2000s only very small single crystals were available just enabling thermodynamic and transport measurements. Then, with substantially improved crystal growth techniques \cite{Seiro2010,Chongde2011}, quite large single crystals of CeCu$_2$Si$_2$ and CeCu$_2$(Si$_{1-x}$Ge$_x$)$_2$ (up to $\sim$\,cm$^3$ size) could be synthesized. Now performing single crystal neutron diffraction on CeCu$_2$(Si$_{1-x}$Ge$_x$)$_2$ the antiferromagnetic order could be followed to much lower Ge concentrations \cite{Stockert2003,Stockert2005}. Finally, incommensurate antiferromagnetic order was even detected in pure $A$-type CeCu$_2$Si$_2$ with a small ordered magnetic moment $\approx 0.1\,\mu_{\rm B}$/Ce \cite{Stockert2004} as shown in Fig. \ref{Fig3_1}(a). The propagation wave vector $\mathbf{k}=\mathbf{Q_{AF}}=(0.215~0.215~0.53)$ at $T=50$\,mK agrees well with theoretical calculations of the Fermi surface using a renormalized band method \cite{Zwicknagl1992,Zwicknagl1993,Stockert2004}. They indicate nesting properties in the corrugated part of the cylindrical Fermi surface of the heavy quasiparticles at the $X$ point of the bulk Brillouin zone [see Fig. \ref{Fig3_1}(b) and Sec.~\ref{ARPESsec}, below]. Hence, the magnetic order in CeCu$_2$Si$_2$ is an incommensurate SDW. This is further supported by the temperature dependence of the propagation wave vector below the ordering temperature. It is worth noting that the propagation vectors in CeCu$_2$(Si$_{1-x}$Ge$_x$)$_2$ are quite similar with the largest difference being the $a*, b*$ component changing from $0.215$ in pure CeCu$_2$Si$_2$ to $0.282$ in CeCu$_2$Ge$_2$ and almost no change in the $c*$ component remaining close to $0.5$ \cite{Stockert2005}. 

The interplay between antiferromagnetism and superconductivity has been studied on small $A/S$-type CeCu$_2$Si$_2$ single crystals, where $\mu$SR measurements indicated a competition of both phenomena with a full repulsion of antiferromagnetism in the superconducting state \cite{Luke1994,Feyerherm1997,Stockert2006}, in contrast to earlier reports on polycrystalline samples \cite{Uemura1988}. Neutron diffraction on quite large $A/S$-type CeCu$_2$Si$_2$ single crystals also revealed that magnetic order and superconductivity do not coexist in CeCu$_2$Si$_2$ on a microscopic scale \cite{Thalmeier2005,Arndt2010}. 

\subsection{Quantum criticality}
\label{QCPSec}

Common to many magnetically ordered Ce-based heavy-fermion systems, the application of pressure tunes the relative strengths of the magnetic exchange interactions (Ruderman-Kittel-Kasuya-Yosida interaction) and Kondo coupling, and for sufficiently large pressures the Kondo interaction dominates, suppressing magnetic order. In several cases this allows for the tuning of a second-order antiferromagnetic transition continuously to zero temperature at a QCP, leading to the breakdown of Fermi liquid behavior at finite temperatures \cite{Stewart2001,Stewart2006,Lohneysen2007,Sachdev2011}. The RKKY interaction leads to antiferromagnetic correlations between the local moments, which reduce the amplitude of the Kondo singlet in the ground state. 

\begin{figure*}[t]
	\begin{center}
		\includegraphics[width=0.99\textwidth]{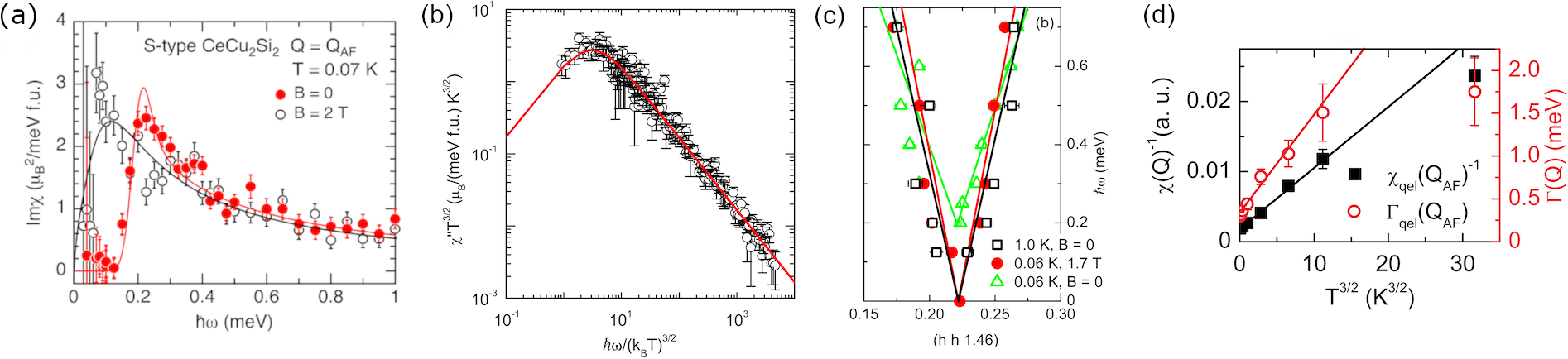}
	\end{center}
	\caption{Spin dynamics in CeCu$_2$Si$_2$. (a) Low-energy spin excitations in $S$-type CeCu$_2$Si$_2$ at ${\mathbf Q}_{AF}$ and $T = 0.07$\, K in the superconducting ($B = 0$) and the normal state ($B = 2$\,T). Reproduced  from \onlinecite{Stockert2011}. (b) Scaling of the normal-state quasielastic response in $S$-type CeCu$_2$Si$_2$ at ${\mathbf Q}_{AF}$ and at $B = B_{c2} = 1.7$\,T indicating universal scaling of the dynamical susceptibility $\chi^{\prime\prime} T^{3/2}$ vs. $\omega/T^{3/2}$. (c) Dispersion of the spin excitations in the normal and superconducting states of $S$-type CeCu$_2$Si$_2$. (d) Relaxation rate $\Gamma$ and inverse susceptibility $\chi(\textbf{Q})^{-1}$ of the normal-state magnetic response at ${\mathbf Q}={\mathbf Q}_{AF}$in $S$-type CeCu$_2$Si$_2$ versus $T^{3/2}$. (b)-(d) are reproduced with permission from \onlinecite{Arndt2011}.}
	\label{Fig3b}
\end{figure*}

Two classes of QCPs have been advanced in recent years, depending on whether this static Kondo-singlet amplitude is destroyed \cite{Si01.1,Col01.1,Senthil2004} or remains nonzero 
 at the antiferromagnetic QCP \cite{Col05.1,Si2010}. Prototype examples of the former case of Kondo-destruction quantum criticality include Au-doped CeCu$_6$ \cite{Schroder2000}, YbRh$_2$Si$_2$ \cite{Paschen2004,Gegenwart2007,Friedemann2010} and CeRhIn$_5$ \cite{Shishido2005,Park2006}. For paramagnetic CeRhIn$_5$ ($p>p_c$), the quantum critical behavior changes at a certain crossover energy scale $E^{*}=k_BT^{*}$ \cite{Park2011b}, suggesting  that the critical fluctuations of the Kondo effect, i.e., partial Mott physics, may be dominating above the crossover scale. CeCu$_2$Si$_2$ shows evidence for a line of QCPs as a function of magnetic field under pressure, in the vicinity of the disappearance of magnetic order, see Fig.~\ref{Fig3}(a). For an $S$-type polycrystalline sample in the low temperature normal state, the signatures of a 3D SDW-type QCP are found from a $T^{\frac{3}{2}}$ dependence of the resistivity, as well as a  $-T^{\frac{1}{2}}$ dependence of the specific heat coefficient \cite{Gegenwart1998}. In addition, the spin-excitation spectrum at the nesting wave vector $\boldsymbol\tau\approx{\mathbf Q}_{AF}$ in the normal state of superconducting ($S$-type) CeCu$_2$Si$_2$ displays an almost critical slowing down when superconductivity is suppressed by a magnetic field \cite{Arndt2011}, as expected for a compound located very close to a QCP. Moreover, an $E/T^{3/2}$ scaling of the normal state magnetic response [Fig.~\ref{Fig3b}(b)] and a $T^{3/2}$ dependence of the inverse lifetime of the spin fluctuations [Fig.~\ref{Fig3b}(d)] indicate that in CeCu$_2$Si$_2$ a 3D SDW-type QCP seems to be realized, in line with the aforementioned thermodynamic and transport measurements \cite{Stockert2011,Arndt2011}. Measurements of the damping rate from inelastic neutron scattering (INS) have provided evidence that the Kondo-destruction temperature scale, $T^{*}$, is nonzero but small \cite{Smidman2018} compared to the bare Kondo scale of $15$~K. Changes in $C(T)/T$ from a square-root to logarithmic dependence and in the quasielastic neutron-scattering damping rate from $T^{3/2}$ to $T$-linear behavior are observed between 1 and 2 K, suggesting that $T^{*}$ is of a similar size.
 
It is to be noted that ${\mathbf Q}_{AF}$ is not a singular point in (${\mathbf Q}$,$\omega$) space, but paramagnons are emerging out of ${\mathbf Q}_{AF}$ with an initial linear dispersion \cite{Stockert2011,Arndt2011}. Comparing the magnetic response in the normal state of superconducting ($S$-type) CeCu$_2$Si$_2$ and  the antiferromagnetic state in $A$-type CeCu$_2$Si$_2$, the dispersion of the (para)magnons in both states was
 found to be very similar, with just higher intensity for the $A$-type sample \cite{Huesges2018}. Upon entering the superconducting state, the dispersion of the paramagnons remains (almost) unchanged with deviations only occurring at low energy transfers below $0.5$\,meV due to the formation of a spin gap \cite{Stockert2011}, see Sec.~\ref{SpinDynSec}. Recently, INS experiments on $S$-type CeCu$_2$Si$_2$ have been extended to higher energy transfers up to several meV \cite{Song2021}. These measurements fully confirm the previous experiments at low energies, i.e., the spin gap in the superconducting state \cite{Stockert2011} and the dispersive paramagnons \cite{Stockert2011,Arndt2011,Huesges2018}. However, in addition, the dispersive spin excitations are now found to change to a dispersionless column in energy above $\approx 1.5$\,meV \cite{Song2021}. The transition from dispersive to dispersionless magnetic excitations occurs around $k_{\rm B}T_{\rm K}$, i.e., the characteristic local energy scale in CeCu$_2$Si$_2$. Currently it is an open question, if and how these high-energy spin excitations are related to the unconventional heavy-fermion superconductivity in CeCu$_2$Si$_2$. 
 
Another issue that has to be clarified by future work concerns the difference in the quantum critical exponent $\alpha$ of the temperature dependence in the low-$T$ resistivity of undoped CeCu$_2$Si$_2$, $\Delta\rho(T) = A'T^{\alpha}$. As already mentioned, this was found to be $\alpha = \frac{3}{2}$ in \cite{Gegenwart1998}, whereas $\alpha=1$ was reported in \cite{Yuan2003,Yuan2006}. In both cases, the samples had been prepared with some Cu excess pointing to $S$-type samples. As shown by neutron diffraction \cite{Stockert2004} as well as earlier $\mu$SR results \cite{Luke1994,Feyerherm1997}, these samples contain a minority phase of $A$-type that is microscopically separated from the $S$-type majority phase. It may be possible that, depending on the content and spatial distribution of this minority phase, the volume-integrated response in resistivity experiments is eventually responsible for the  differing $T$-dependences observed.

\subsection{Spin dynamics in superconducting C\lowercase{e}C\lowercase{u}$_2$S\lowercase{i}$_2$}
\label{SpinDynSec}

Due to the small magnetic moment and the low transition temperatures, INS experiments were performed on superconducting CeCu$_2$Si$_2$ using cold-neutron triple-axis spectroscopy. While the INS spectra in the normal state yield a quasielastic magnetic response at ${\mathbf Q}_{AF}$ with slowing down and scaling behavior as already mentioned, the spin dynamics in the superconducting state well below $T_c = 0.6$\,K show a clear spin excitation gap at ${\mathbf Q}_{AF}$\cite{Stockert2008,Stockert2011} followed by a well defined maximum, often called `spin resonance' [Figs.\ref{Fig3b}(a) and (b)]. It should be noted that this maximum clearly exceeds the magnetic response
in the normal state, in contrast to a simple $s$-wave superconductor where no obvious enhancement of the superconducting response over the normal state response is expected at energies above the spin gap. Its intensity depends on the Fermi surface topology and the paramagnon dispersion \cite{Eremin2008} and might therefore be less pronounced than in other unconventional superconductors. With a spin-gap size of about 0.2 meV, the maximum is located at $4 k_{\rm B}T_c$ and its position is therefore smaller than $2\Delta=5k_{\rm B}T_c$ of the (large) charge gap \cite{Fujiwara2008}. We note that this necessary condition for a 'spin resonance' to be located inside  $2\Delta$ was also fulfilled by the low-energy peak in UPd$_2$Al$_3$ (in which the U$^{3+}$ ion has two localized and one more hybridized $5f$-electron), where heavy-fermion superconductivity coexists with local-moment antiferromagnetic  order \cite{Sato2001}. Like in the latter case as well as in CeCoIn$_5$ \cite{Song2016,Song2020}, the peak in CeCu$_2$Si$_2$ develops in the one-particle channel, i.e. out of the aforementioned quasielastic line that persists  to way above $T_c$ for the two Ce-based compounds, and to well above $T_N$ ($>T_c$) for UPd$_2$Al$_3$. This is different from the cuprates where it manifests a singlet-triplet excitation of the $d$-wave condensate \cite{Sidis2004}.

Though this distinct maximum in the INS data at the edge of the spin-excitation gap should not be called a 'spin resonance' for the reasons given above, it nevertheless highlights a sign-changing superconducting order parameter. Namely, if one considers coupling between a magnetic mode (e.g. magnon/magnetic exciton) and the itinerant quasiparticles/Cooper pairs \cite{Bernhoeft1998, Bernhoeft2006}, the observation in CeCu$_2$Si$_2$ of a significant low energy enhancement of the INS intensity along the propagation vector $\mathbf{Q}_{\rm AF}$ in the superconducting state over that of the normal state  \cite{Stockert2011}, implies a large coherence factor , which 
 necessarily requires a sign change of the superconducting order parameter along this wave vector. Alternatively, for CeCoIn$_5$ and Fe-based superconductors it has been proposed that the low energy INS peak arises from reduced quasiparticle damping in the superconducting state, allowing for the observation of an otherwise overdamped magnon mode \cite{Chubukov2008,Onari2010}.  For two reasons, we do not consider this scenario to be viable.  First, the ratio of the energy of the INS maximum to $2\Delta$ is comparable to the universal value observed in a variety of correlated superconductors \cite{Yu2009,Duan2021}. Second (and related to the first), for this scenario to occur, the universality of the INS peak in the superconducting
state which occurs in a variety of systems requires some degree of commonality in the behavior of the  low-energy spin excitations in their normal states. This expectation is to be contrasted with disparate behavior of the low-energy spin excitations that have been observed in the normal state of these systems. In particular, in the case of CeCu$_2$Si$_2$, even in the normal state the paramagnons do not appear to be overdamped as suggested by their well visible dispersion at low energies [see Fig.~\ref{Fig3b}(c)], even up to $k_BT_K\approx1.5$~meV \cite{Song2021}.

The  experimentally determined propagation vector $\mathbf{Q}_{\rm AF}$ agrees very well with the theoretically obtained nesting wave vector $\boldsymbol\tau$ shown in Fig.~\ref{Fig3_1}(b) \cite{Zwicknagl1992,Zwicknagl1993, Stockert2004}, which connects nested parts of the heavy quasiparticle band, highlighting \textit{intraband} nesting. Importantly,  $\mathbf{Q}_{\rm AF}$ does not connect extended regions of different bands (interband nesting) of e.g. electron and hole bands (Sec.~\ref{ARPESsec}) as  required by the $s_{+-}$ pairing model that has been considered for some Fe-based superconductors \cite{Mazin2008}.

\subsection{Effects of potential scattering}
\label{impSec}

\begin{figure}[t]
	\begin{center}
		\includegraphics[width=0.6\columnwidth]{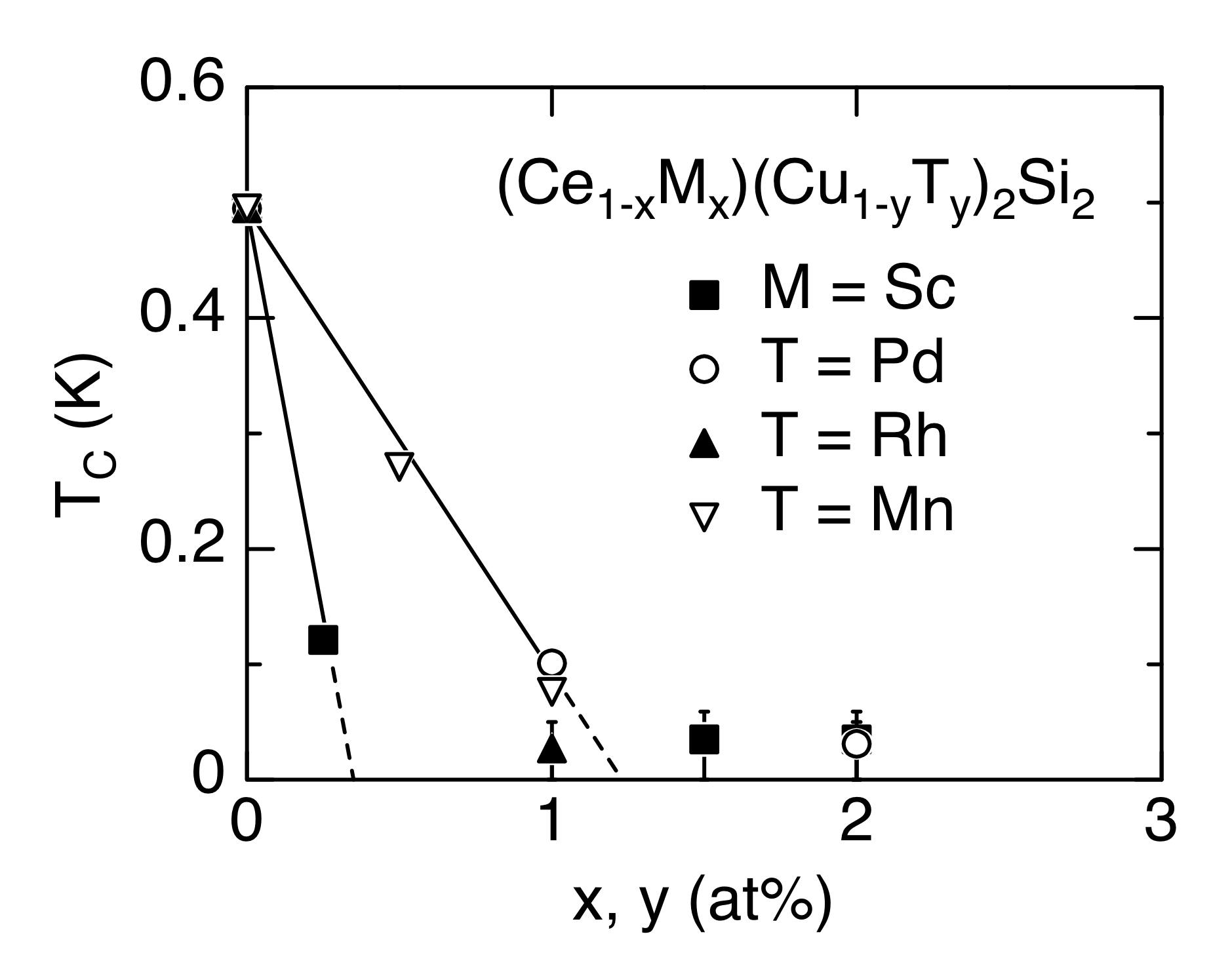}
	\end{center}
	\caption{Dependence of the superconducting transition temperature of CeCu$_2$Si$_2$ polycrystals on substitutions for Ce and Cu. Replotted from \cite{Spille1983}.}
	\label{Fig10}
\end{figure}

Historically, the effect of nonmagnetic impurities has been an important test for unconventional superconductivity. This is because, while the $T_{\rm c}$ of a conventional BCS superconductor is very sensitive to magnetic impurities, the non-magnetic case has little effect \cite{Anderson1959}. On the other hand, for superconductors with unconventional sign-changing states, the effect of nonmagnetic impurities may become similar to magnetic impurities in a conventional material \cite{Balatsky2006}. Indeed the high sensitivity of CeCu$_2$Si$_2$ to a very small amount of atomic substitution of nonmagnetic impurities was one of the key pieces of early evidence allowing for the identification of an unconventional superconducting state. This is particularly the case for substitutions on the Cu site, where as shown in Fig.~\ref{Fig10}, doping around 1\% of Rh, Pd or Mn completely suppresses $T_{\rm c}$, while similarly only 0.5\% of smaller Sc$^{3+}$  on the Ce site is needed \cite{Spille1983}. A striking difference for the Ce site is the size dependence of the dopant, where $T_{\rm c}$ becomes increasingly insensitive for larger substituents, i.e., with critical concentrations of 6\% for Y$^{3+}$, 10\% for La$^{3+}$  \cite{Spille1983}, culminating in 20\% for Th$^{4+}$ \cite{Ahlheim1992}.  This  trend with chemical pressure is analoguous to that found when applying hydrostatic pressure to CeCu$_2$Si$_2$ doped with 10 at\% of Ge, where $T_{\rm c}$ is suppressed on the high pressure side of the low-pressure dome centered around the antiferromagnetic QCP  [Fig.~\ref{twodome}]. While this size effect appears to be in line with the strength of the Kondo interaction in the dependence of the volume available to the Ce$^{3+}$ ions, the reason for the distinct site dependence in the atomic substitution experiments is yet to be unraveled. 

Such small critical substitutions on the transition-metal side being needed to suppress $T_{\rm c}$  proves that this cannot be simply due to a significant tuning of the Kondo state. Indeed, while Ge doping expands the lattice acting as a negative pressure effect and causes a slight decrease of $T_K$, it is found that tuning a Ge-doped sample using pressure, which causes an increase of $T_K$, still yields a greatly suppressed $T_{\rm c}$ \cite{Yuan2003,Yuan2006}. Such a reduction of $T_{\rm c}$ upon 10\% Ge-doping allowed for the revelation that there are two separate superconducting domes in the temperature-pressure phase diagram, one sitting near a magnetic QCP, while the higher pressure dome potentially lies near a valence transition (Fig.~\ref{twodome}) \cite{Yuan2003,Holmes2004}. Meanwhile for a more disordered sample with 25\% Ge-doping no superconductivity is recovered even after the suppression of magnetism by pressure \cite{Yuan2004}.

On the basis of recent studies of electron irradiated samples, it was proposed that the order parameter of CeCu$_2$Si$_2$ does not change sign across the Fermi surface, much like a conventional BCS superconductor. Namely it is reported that the suppression of $T_{\rm c}$ upon the introduction of disorder by electron irradiation is not as rapid as expected for sign-changing pairing states such as those in the cuprates or iron pnictides, but instead is similar to some materials believed  to have a conventional pairing mechanism \cite{Yamashita2017}. Moreover the lack of change in the low-temperature penetration depth of electron irradiated samples  is taken as evidence for a lack of low-energy impurity-induced bound states, as also expected for sign-preserving order parameters \cite{Takenaka2017}. Since the effect of electron irradiation is likely to correspond to the displacement of Ce atoms from the lattice to interstitial sites, the resulting disorder may be compared to that manifested by the strong (factor of four) variation in the residual resistivity $\rho_0$ going from a nearly stoichiometric $A/S$-type to an $S$-type single crystal with a small amount of Cu excess, where no depression of $T_{\rm c}$  is observed \cite{Pang2018}. Similar results are well known from the cuprate high-$T_{\rm c}$ superconductors where substantial variations in $\rho_0$ are not reflected by any significant changes in $T_{\rm c}$, cf. results on YBCO polycrystals \cite{Cava1987} and single crystals \cite{Liang1992}. As discussed in Sec.~\ref{Theorsub}, there are a number of theoretical works underlining the robustness of unconventional superconductivity against certain kinds of ordinary potential scattering \cite{Si2016,Anderson1997}.
However, from the cuprates it is also known that atomic substitution can be quite hostile for high-$T_{\rm c}$ superconductivity \cite{Alloul2009}. For example, partial substitution of Cu on the CuO$_2$ planes by Zn causes a strong depression of $T_{\rm c}$  \cite{Xiao1998}. Obviously, this is quite similar to the results of the aforementioned substitution experiments on CeCu$_2$Si$_2$ \cite{Spille1983,Ahlheim1988,Yuan2003}, which are at odds with a non-sign changing superconducting state. 

In summary, the aforementioned studies on CeCu$_2$Si$_2$ reveal that ``impurity doping", i.e., substitutional  disorder, is strongly pairbreaking, while certain kinds of lattice rearrangements, induced, e.g., by electron irradiation or small changes in the Cu/Si occupation, are harmless to superconductivity. This dichotomy of harmful and harmless disorder in unconventional heavy fermion and cuprate high-$T_{\rm c}$ conductors still needs to be uncovered.

\subsection{Evidence for $d$-wave pairing}
\label{dwavesec}

\begin{figure}[t]
	\begin{center}
		\includegraphics[width=0.99\columnwidth]{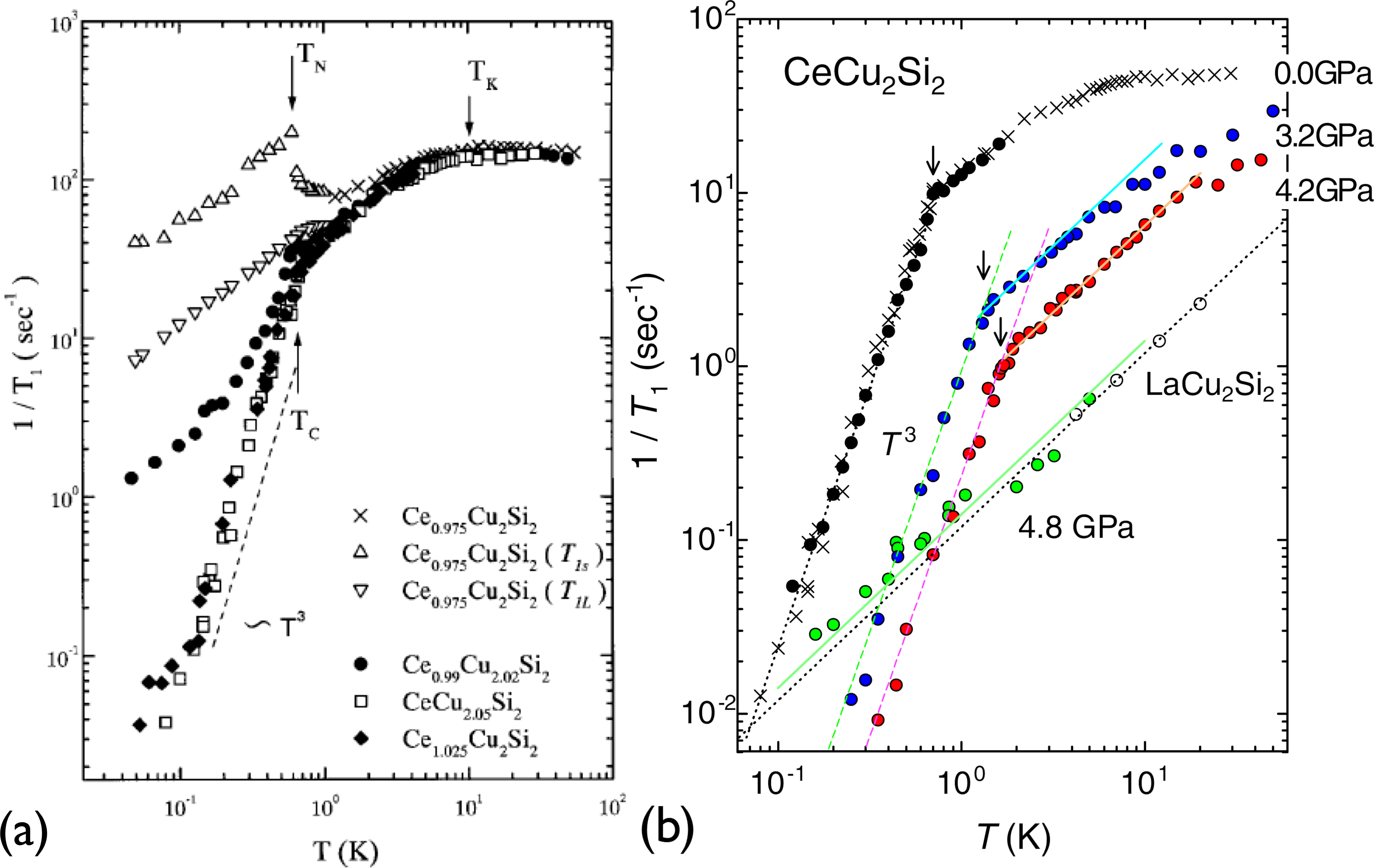}
	\end{center}
	\caption{Temperature dependence of the spin lattice relaxation rate  $1/T_1(T)$ of CeCu$_2$Si$_2$, obtained from Cu-NQR measurements (left) for various superconducting and non-superconducting polycrystals,  and (right) single crystals of superconducting CeCu$_2$Si$_2$ under hydrostatic pressure, as well as LaCu$_2$Si$_2$. No Hebel-Slichter peak at $T_{\rm c}$ is observed and a  $T^3$ dependence is found in the superconducting samples down to around 0.1~K. The left panel is reproduced with permission from \onlinecite{Ishida1999}. The right panel is reproduced from \onlinecite{Fujiwara2008}, copyright 2008 The Physical Society of Japan.}
	\label{Fig3a}
\end{figure}

For a long time, the pairing state of CeCu$_2$Si$_2$ was generally believed to correspond to $d$-wave superconductivity, in line with other Ce-based heavy-fermion superconductors \cite{Thompson2012}, cuprate materials \cite{Scalapino1995,Lee2006}, and organic superconductors \cite{Kanoda2008,Lang2003}. A decrease of the Knight shift below $T_{\rm c}$, the ordinary size of the dc Josephson effect between polycrystalline CeCu$_2$Si$_2$ and Al as well as evidence for Pauli limiting of the upper critical field (Fig.~\ref{Fig2}), confirmed quite early that the Cooper pairs correspond to a singlet pairing state \cite{Ueda1987,Assmus1984}. Meanwhile the clearest evidence for the superconducting gap structure came from Cu-NQR measurements, where the spin lattice relaxation rate ($1/T_1(T)$) displayed in Fig.~\ref{Fig3a} shows a $T^3$ dependence down to around 0.1~K, which is characteristic of line nodes in the superconducting gap \cite{Ishida1999,Fujiwara2008}, although it should be noted that the $1/T_1(T)$ data of Ishida \textit{et al.} \cite{Ishida1999} also show some deviation from $T^3$ behavior at the lowest temperatures, as clearly demonstrated in recent NQR experiments extended to somewhat lower temperature \cite{Kitagawa2017}, see below. Evidence for nodal superconductivity was also inferred from measurements of other thermodynamic quantities, including a $T^2$ dependence of the magnetic penetration depth  \cite{Gross1998}, which  is consistent with $d$-wave superconductivity in the presence of strong impurity scattering \cite{Hirschfeld1993}. The  requirement that the order parameter is (i) spin singlet, (ii) with gap nodes, and (iii) changes sign on the regions of the renormalized Fermi surface connected by the nesting wave vector $\boldsymbol\tau\approx{\mathbf Q}_{AF}$, is most readily satisfied by a $d_{x^2-y^2}$ pairing state, similar to that generally believed to apply to the cuprate high-$T_{\rm c}$ superconductors \cite{Scalapino1995}. On the other hand, in isothermal magnetoresistance measurements the angular dependence of the upper critical field in the $ab$-plane at 40~mK was found to be most compatible with a $d_{xy}$ state, although the small amplitude of this modulation made it difficult for firm conclusions to be drawn \cite{Vieyra2011}. Nevertheless, a $d$-wave pairing state of some form with line nodes was long considered to be the most likely candidate pairing state.

\section{Fully gapped unconventional superconductivity in C\lowercase{e}C\lowercase{u}$_2$S\lowercase{i}$_2$}

\subsection{Evidence for a nodeless gap structure}
\label{FullgapSec}

\begin{figure}[t]
	\begin{center}
		\includegraphics[width=0.95\columnwidth]{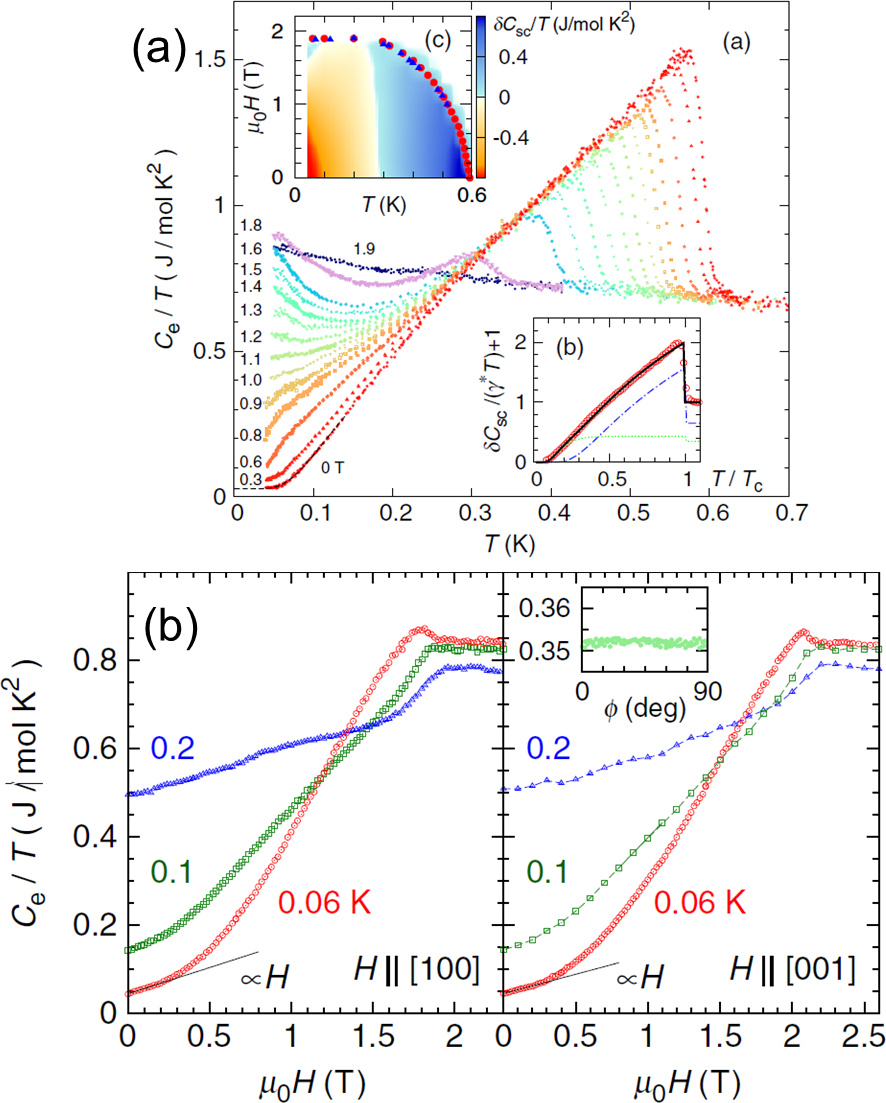}
	\end{center}
	\caption{(a) Temperature dependence of the electronic contribution to the specific heat as  $C_{\rm e}/T$ of CeCu$_2$Si$_2$ down to temperatures of 0.04~K. The lower inset displays the analysis of the data using a nodeless two-gap model, while the upper panel shows the data as a contour plot. (b) The field dependence of the electronic specific heat coefficient  at various temperatures for fields along the [100] (left) and [001] directions. At 0.06~K, linear behavior is observed at low fields for both field orientations. The inset in the right panel shows $C_{\rm e}/T$ as a function of the in-plane azimuthal field-angle $\phi$, which remains constant. Reproduced with permission from \onlinecite{Kittaka2014}. }
	\label{Fig4}
\end{figure}	

The understanding of the superconducting state of CeCu$_2$Si$_2$ underwent a radical overhaul following the results from low-temperature specific heat measurements of Kittaka \textit{et al.} \cite{Kittaka2014,Kittaka2016}, which revealed that the superconducting gap is fully open over the whole Fermi surface. Here the temperature dependence of the electronic contribution to the specific heat $C_{\rm e}$ ($\approx C_{4f}$) of  $S$-type single crystals measured down to 0.04~K begins to flatten upon approaching the lowest measured temperature, and was best described by an exponentially activated temperature dependence, rather than following the $C_{\rm e}\sim T^2$ behavior of a superconductor with line nodes, as shown in Fig.~\ref{Fig4}(a). This analysis suggested nodeless superconductivity with a gap $\Delta_0=0.39k_{\rm B}T_{\rm c}$. Since this is considerably less than the value of $1.76k_{\rm B}T_{\rm c}$ derived from weak-coupling BCS theory, in order to demonstrate the presence of a fully open gap in thermodynamic quantities such as the specific heat and penetration depth, measurements across a wide temperature range  down to at least 0.05~K are required. A further advantage of this  study is the very small residual $\gamma_0=0.028$~J~mol$^{-1}$~K$^{-2}$, which again allows for the inference of a lack of low-energy excitations. After subtracting an estimate of the phonon contribution, the data up to $T_{\rm c}$ could not be described by a model with a single gap, but were instead accounted for by a model with two nodeless isotropic gaps. 

These conclusions were supported by specific heat measurements in applied magnetic fields, displayed in Fig.~\ref{Fig4}(b). Here the isothermal $C_{\rm e}/T$ at the lowest temperature of  0.06~K exhibits a linear field dependence, as opposed to the $H^{0.5}$ behavior of a $d$-wave superconductor. The range of this low-field linear region is relatively narrow, and at higher fields there is a pronounced increase of $C_{\rm e}/T$. Just below $B_{\rm c2}$, $C_{\rm e}/T$  even overshoots the normal state value, and the origin of this strong enhancement needs still to be clarified by future work. Upon rotating the field within the $ab$-plane, no modulation of the specific heat is observed, whereas in the single band $d$-wave scenario a four-fold oscillation is predicted theoretically \cite{Vorontsov2007,Boyd2009}, and observed experimentally in the Ce$T$In$_5$ series of heavy-fermion superconductors \cite{Aoki2004,An2010,Lu2012}. Furthermore, measurements as a function of the polar angle $\theta$ reveal simply the two-fold oscillations arising naturally from the tetragonal symmetry \cite{Kittaka2016}.

 It should be noted that early evidence for a potentially exponential temperature dependence of the low-temperature specific heat was provided by measurements of CeCu$_2$Si$_2$ polycrystals, which revealed power-law behaviour with an exponent of two near $T_{\rm c}$, but close to three at $T =~ 0.05$~K \cite{Steglich1985a}. Further early evidence for fully gapped superconductivity was reported from a point contact study of CeCu$_2$Si$_2$ measured at 0.03~K \cite{Wilde1994}. They found that the differential resistance curves are flat around zero bias, which is characteristic of a fully open gap,  in stark contrast to that observed in UPt$_3$ where the curves have a triangular shape around zero voltage suggesting the presence of gap nodes.

\begin{figure}[t]
	\begin{center}
		\includegraphics[width=0.8\columnwidth]{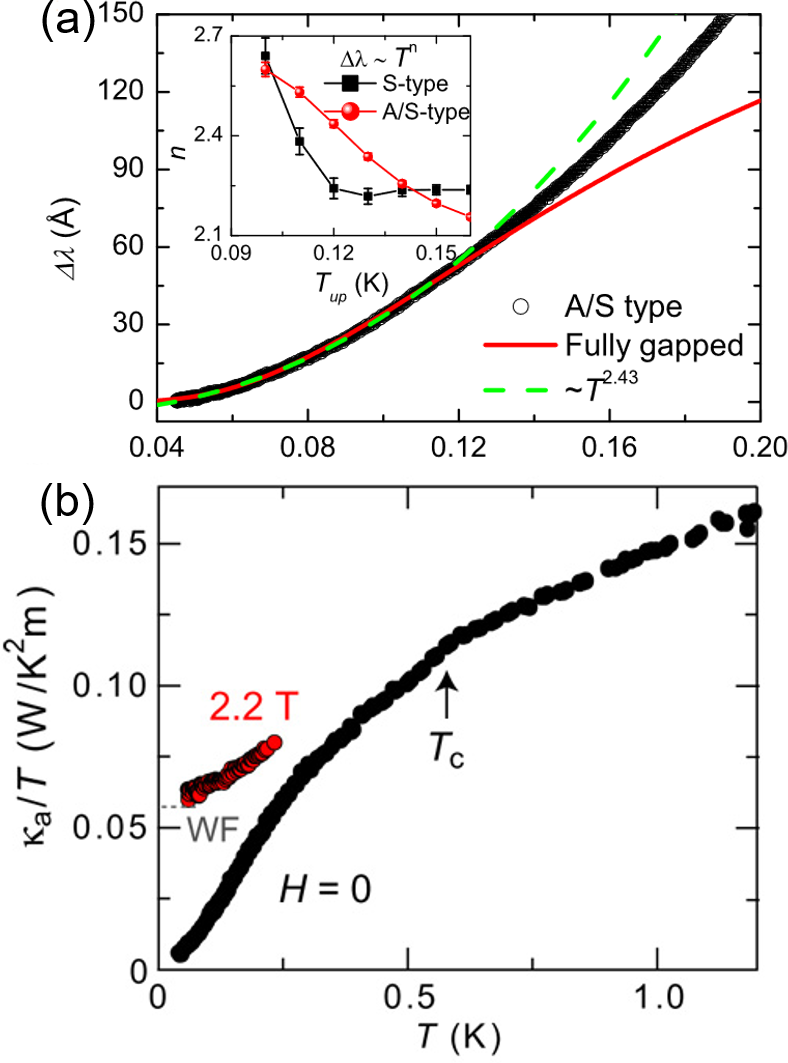}
	\end{center}
	\caption{(a) Temperature dependence of the magnetic penetration depth shift of an $A/S$-type CeCu$_2$Si$_2$ single crystal,  measured using the tunnel-diode-oscillator-based method. The solid line shows the fit to the low temperature data with  a model for a fully gapped superconductor, where the lack of nodes in the gap is corroborated by an exponent $n$ that is consistently larger than two (inset). Reproduced from  \onlinecite{Pang2018} (b) Temperature dependence of the in-plane thermal conductivity as $\kappa_{\rm a}/T$  of CeCu$_2$Si$_2$, which in the superconducting state ($H=0$) extrapolates to zero at $T=0$, clearly demonstrating fully gapped superconductivity. Also shown are the data in the normal state, which demonstrates the validity of the Wiedemann-Franz law. Reproduced from \onlinecite{Yamashita2017}, available under a Creative Commons NonCommercial 4.0 International Public License. }
	\label{FGFig}
\end{figure}

Penetration depth measurements performed down to $T\approx0.05$~K also demonstrate a fully open gap \cite{Yamashita2017,Takenaka2017,Pang2018}. As shown in Fig.~\ref{FGFig}(a), the  low temperature penetration depth shift $\Delta\lambda(T)$ is well described by the expression for a fully gapped material, with gap values well below that of BCS theory (in the range of $0.5-1k_{\rm B}T_{\rm c}$). Moreover, when analyzed using a power law dependence $\Delta\lambda(T)\sim T^n$ in temperature intervals with decreasing upper limits, the low temperature exponents are consistently found to increase with $n>2$, exceeding the bounds expected for a line nodal superconductor of $n=1$ and $n=2$ in the clean and dirty limits, respectively. 

Fully gapped superconductivity was also deduced from recent thermal conductivity measurements, where the coefficient of the in-plane thermal conductivity $\kappa_a/T$ extrapolates to zero at zero temperature [Fig.~\ref{FGFig}(b)], again showing evidence for the  lack of low-energy excitations expected for nodeless superconductivity \cite{Yamashita2017}. This is further supported by  measurements of the magnetic-field dependence of  $\kappa_a/T$ , where there is little change with applied field in the low field region. It is noted that more ambiguous results were found from earlier thermal conductivity studies of CeCu$_2$Si$_2$ \cite{VieyraThesis}, while the recent measurements reported by \cite{Yamashita2017} benefited from samples with  lower non-superconducting fractions, as well as better contacts between the heater and the sample. Whereas earlier NQR measurements were found to exhibit  a $T^3$ dependence of $1/T_1(T)$ down to around 0.1~K \cite{Ishida1999,Fujiwara2008}, more recent results show a deviation from this behavior at very low temperatures which can be accounted for by a small but nodeless gap \cite{Kitagawa2017}.  

While most recent low-temperature measurements have indicated that the superconducting gap is fully open, including also small-angle neutron scattering measurements \cite{Campillo2021}, results from a low-temperature scanning spectroscopy study were less conclusive, which may be related to the fact that no good cleaves have been achieved in CeCu$_2$Si$_2$ until very recently (see  Sec.~\ref{ARPESsec}). The tunneling spectra measured at low temperatures show two clear features at different voltage bias, providing clear evidence for multiple gaps or an anisotropic gap structure \cite{Enayat2016}. However, the data were best accounted for by a model where the large gap is fully open, but the small gap is nodal. The reason for this discrepancy is not clear, but it should be noted that the density of states of the $d+d$ band mixing pairing state (Sec.~\ref{Theorsub}) is linear for energies just above the small gap parameter, much like a line nodal superconductor, and therefore this could reconcile these results with other recent findings of nodeless superconductivity.

\subsection{Fermi surface and quasiparticle dispersion}
\label{ARPESsec}

\begin{figure}[t]
\centering
\includegraphics[scale=0.7]{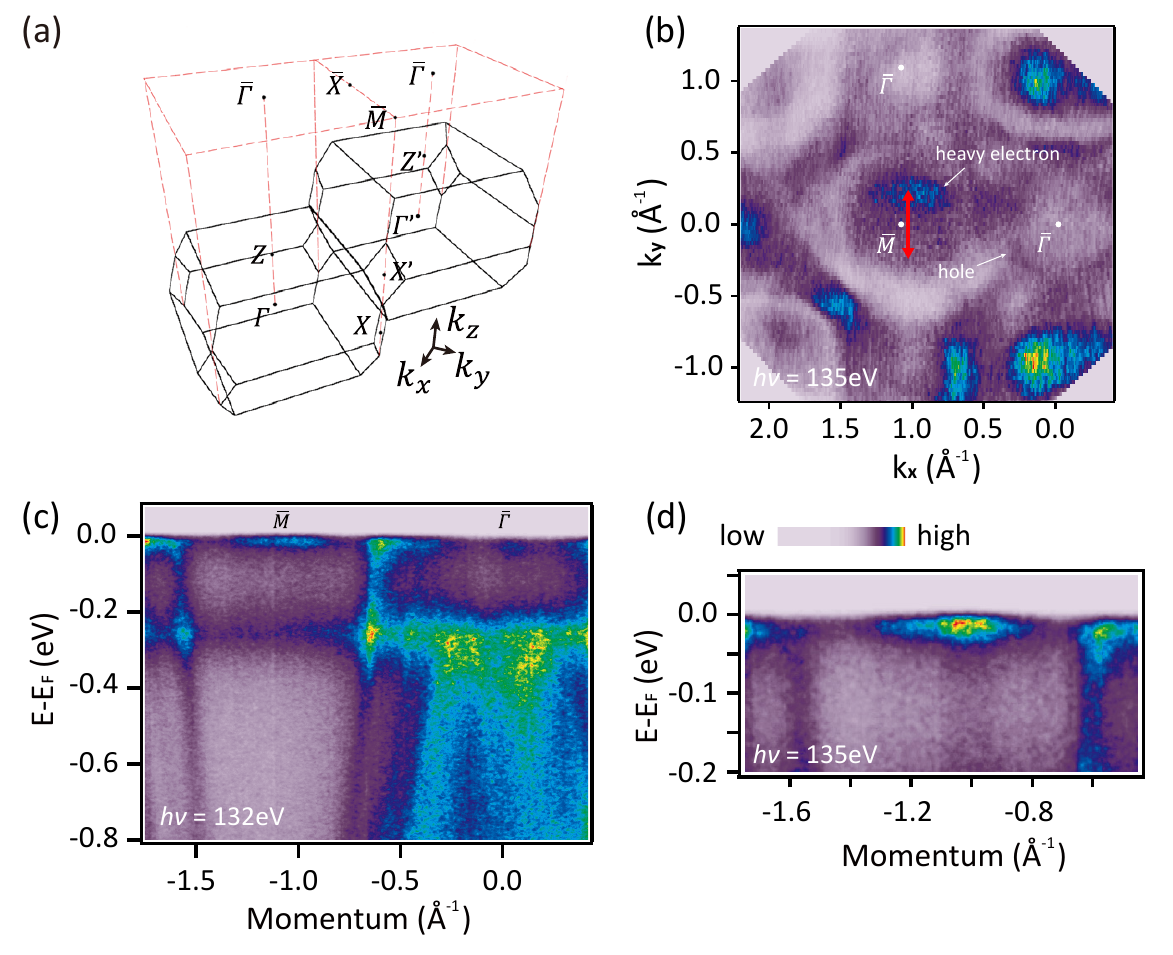}
\caption{The Fermi surface  and quasiparticle dispersion of S-type CeCu$_2$Si$_2$ from ARPES measurements. (a) Three-dimensional bulk Brillouin zone (black solid lines) and the projected surface Brillouin zone (red dashed lines) of CeCu$_2$Si$_2$. (b) The experimental $k_x$-$k_y$  map at 10~K taken with 135 eV photons. The red arrow indicates the in-plane component of the SDW ordering wave vector $\mathbf{Q_{AF}}$ observed by neutron diffraction (Sec.~\ref{AOrigin}). (c) Band dispersion along $\Bar{\varGamma}$-$\bar{M}$ at 10 K. (d) Zoom-in view of the heavy electron band near $E_F$. Reproduced with permission from \onlinecite{Wu2021}.}
\label{Fig7}
\end{figure}

In order to unravel the electronic correlations and superconductivity in CeCu$_2$Si$_2$, the Fermi surface  and quasiparticle dispersions close to $E_F$ are crucial. While the Fermi surface of CeCu$_2$Si$_2$ has been predicted by a number of theoretical studies \cite{Zwicknagl1993,Zwicknagl2016,Ikeda2015,Li2018,Pourovskii2014,Luo2020}, direct momentum-resolved measurements from angle-resolved photoemission spectroscopy (ARPES) are challenging due to the difficulty of sample cleavage \cite{Reinert2001}. Recently, such experimental obstacles have been overcome due to an improved sample preparation method and a newly developed ARPES technique with a small beam spot \cite{Wu2021}. Figure~\ref{Fig7} summarizes the ARPES results from a typical $S$-type single crystal. The experimental Fermi surface  of CeCu$_2$Si$_2$ consists of three-dimensional hole bands centered at the bulk \textit{Z} point [projecting onto the $\Bar{\varGamma}$ point of the surface Brillouin zone] and a quasi-2D electron band at the $X$-point ($\bar{M}$ point at the surface Brillouin zone corner), see Figs.~\ref{Fig7}(a) and (b). Measurements of the energy-momentum dispersion show that the quasi-2D electron band is of predominant $4f$ character and possesses a large effective mass [Figs.~\ref{Fig7}(c) and (d)], while the hole bands near the $\Bar{\varGamma}$ point are mainly derived from the lighter conduction bands.

The heavy electron band observed near the $\bar{M}$ point makes an important contribution to the Fermi surface  and is crucial for the heavy-fermion superconductivity \cite{Zwicknagl1993}. Photon-energy dependent scans and detailed analysis reveal that this heavy electron band is cylindrical in momentum space  and has an effective mass of $\approx$~120$m_e$. Here the effective mass is estimated by first dividing the experimental ARPES spectra by the (resolution-convoluted) Fermi-Dirac distribution function and then fitting the extracted quasiparticle dispersion with a parabola.  Due to the limited energy resolution in ARPES, the effective mass estimation can have relatively large uncertainty. Note that the (zero temperature) effective mass used in the renormalized band calculation is $\approx500m_e$ \cite{Zwicknagl1993,Zwicknagl2016}. Given that the ARPES was performed down to 10~K, at which temperature  $C_{4f}/T$ ($\approx0.125$~J$\cdot$mol$^{-1}\cdot$K$^{-2}$ [Fig.~\ref{Fig1new}]) is approximately seven times smaller than in the low temperature limit \cite{Steglich1990}, the estimated effective masses indicate a very good correspondence between ARPES and the specific heat. As illustrated in Fig.~\ref{Fig3_1}(b), renormalized band calculations \cite{Zwicknagl1993,Stockert2004} reveal that this heavy electron band has a warped part with flat parallel sides connected by a nesting vector $\boldsymbol\tau$, in excellent agreement with the SDW ordering wave vector $\mathbf{Q_{\rm AF}}$ observed in neutron diffraction [Figs.~\ref{Fig3_1}(a) and (b)] \cite{Stockert2004}. The experimental contour of this heavy band shown in Fig.~\ref{Fig7}(b) is in fairly good agreement with these calculations.  Another interesting observation is that the outer hole band near the $\Bar{\varGamma}$ point contains appreciable $4f$ weight and bends slightly near $E_F$, which is the hallmark of  hybridization between conduction and $4f$ electrons \cite{Im2008,Chen2017,Jang2020,Wu2021b}. Its enclosed area is close to the values obtained from quantum oscillation  measurements \cite{Hunt1990,Tayama2003}, which however, could not detect the heavy electron band at the $\bar{M}$ point. Note that the detection of heavy bands can be particularly challenging in quantum oscillation experiments, due to the rapid decay of the quantum oscillation amplitudes with temperature for heavy orbits \cite{Shoenberg2009}.

\subsection{$d+d$ matrix-pairing state}
\label{Theorsub}

\begin{figure}[t!]
\includegraphics[width=0.65\columnwidth]{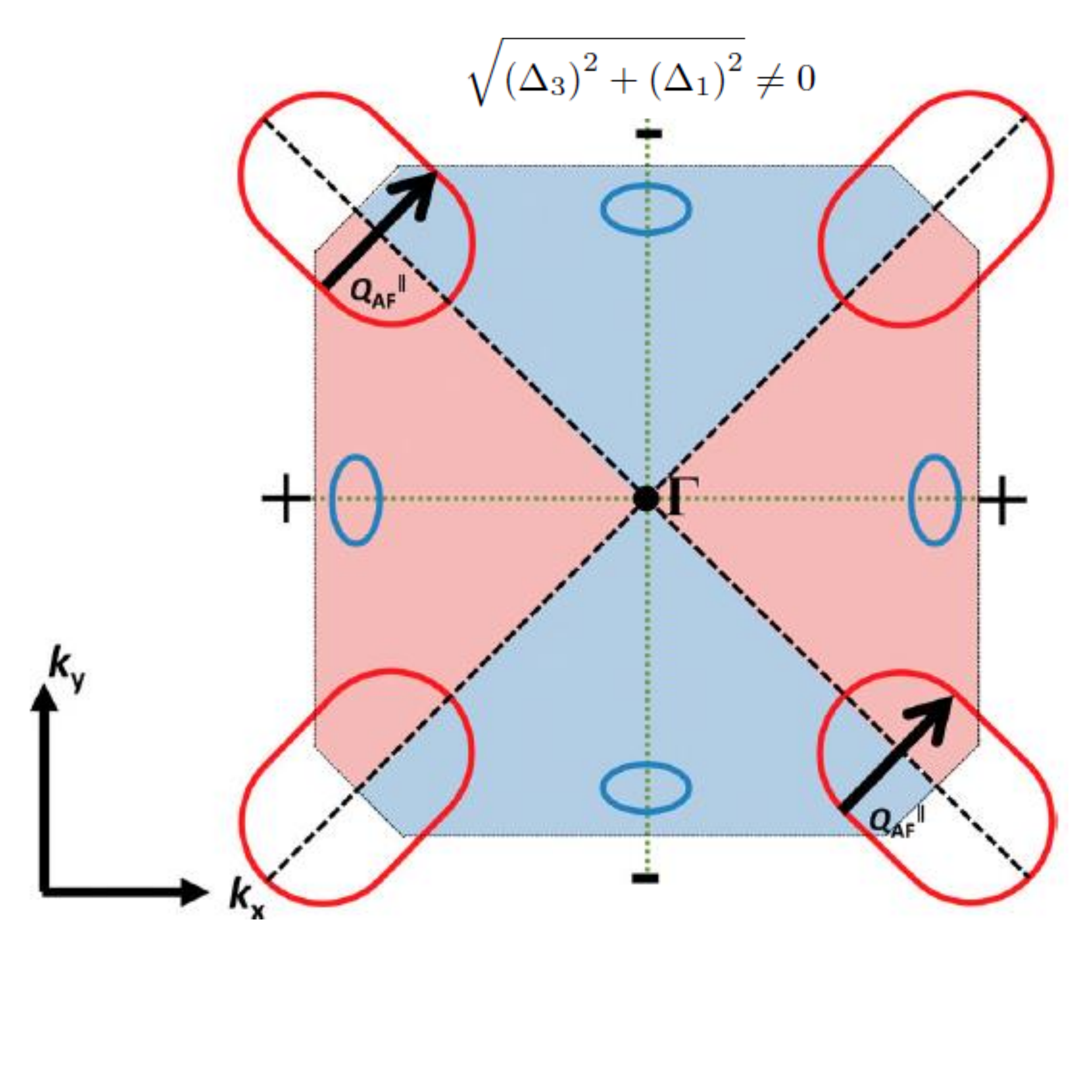}
\caption{Projection of the renormalized heavy Fermi surface  and 
ordering wave vector $\mathbf{Q}_{\text{AF}}=(0.215, 0.215, 0.53) $ 
onto the $k_{z}=0$ plane of the 3D Brillouin zone 
at zero-temperature. 
The dashed lines indicate the nodes of the individual components of $d+d$ pairing. The diagonal black  and vertical/horizontal green dashed lines denote the nodes of $\Delta_{3}(\textbf{k}) \propto d_{x^{2}-y^{2}}$ and $\Delta_{1} (\textbf{k}) \propto d_{xy}$, respectively (see Eq.~\ref{Eq:dpd}). The effective gap is determined by the addition in quadrature of the two components. Since the nodes of the $\Delta_{3}$ and $\Delta_{1}$ components do not overlap except at isolated points of the Brillouin zone, $d+d$ pairing is always gapped on the Fermi surface.
The wave vector for the peak of the observed 
antiferromagnetic fluctuations
projected onto the $(k_x,k_y)$ plane,
${\mathbf Q}_{AF}^{||}$, connects parts of the cylindrical Fermi surface  near the edges 
(red pill-shapes)
where $\Delta_{3} \propto d_{x^{2}-y^{2}}$ 
has opposite signs, leading to the emergence of a pronounced peak inside the superconducting gap in INS experiments.  Reproduced from \onlinecite{Pang2018}.
}
\label{BZ_schm}
\end{figure}

As discussed in Sec.~\ref{dwavesec}, the majority of the experiments in superconducting CeCu$_{2}$Si$_{2}$ have 
been interpreted in terms of a single-band, $d$-wave Cooper pairing. 
The new results presented in Sec.~\ref{FullgapSec} point toward the emergence of a full gap. 
Although a single-band $d$-wave pairing state is at odds with the more recent results, 
the underlying sign-changing nature of the pairing state under a $C_{4z}$ rotation continues to play an important role. This is best illustrated by the large peak 
 in the INS intensity in the superconducting state [Fig.~\ref{Fig3b}(a)] associated with a pairing state which changes sign \emph{within} the heavy, cylindrical bands near the edge of the Brillouin zone, as illustrated in Fig.~\ref{BZ_schm}. 

\noindent 

The robustness of the sign-changing nature of the pairing states suggests that new pairing candidates have to reconcile this feature 
with the emergence of a full gap. An important 
requirement
is that the sign-changing
 but also gapped pairing state must belong to a single irreducible representation 
 of the point group. Indeed, unlike systems such as UPt$_3$ \cite{Fisher1989,Schemm2014},  there have been 
no reports of multiple superconducting transitions in CeCu$_2$Si$_2$ which further break symmetry 
with decreasing temperature. Similarly, the lack of evidence for time-reversal 
symmetry breaking in the superconducting state makes $d+id$ or $s+id$ pairing states unlikely, since these are gapped and sign-changing but only at the price of breaking both 
point-group and time-reversal symmetries. We instead consider a pairing state that can 
reconcile the features of superconducting CeCu$_{2}$Si$_{2}$ 
while preserving the symmetries already mentioned.
This is a multi-band $d+d$ pairing of concurrent intra-band $d_{x^{2}-y^{2}}$- 
and 
inter-band $d_{xy}$-waves \cite{Nica2017,Nica2021}.
In it's most general form, $d+d$ pairing is 

\noindent \begin{align}
\Delta_{d+d} = & 
\begin{pmatrix}
\Delta_{3}(\textbf{k}) & \Delta_{1}(\textbf{k}) \\
\Delta_{1}(\textbf{k}) & - \Delta_{3}(\textbf{k})
\end{pmatrix},
\label{Eq:dpd}
\end{align}

\noindent where the intra- and inter-band components $\Delta_{3}$ and $\Delta_{1}$ transform as $d_{x^{2}-y^{2}}$ and $d_{xy}$, respectively.
This \emph{matrix-pairing} state, which is intrinsically multi-band, has additional structure due to the band space 
on which it is defined. The intra-band $d_{x^{2}-y^{2}}$ component naturally 
satisfies the required sign-change, much like a single-band $d$-wave pairing. 
In contrast to the latter, the matrix structure of $d+d$ pairing, due to the anti-commuting Pauli matrices, also ensures that the gap is determined 
by the addition in quadrature of the two distinct $d$-wave components. Consequently, the Bogoliubov-de Gennes (BdG) quasiparticle 
spectrum shows a full gap
everywhere on the Fermi surface. As recently discussed in \cite{Nica2021}, $d+d$ pairing is a natural $d$-wave
 analogue to the spin-triplet pairing states of $^{3}$He-B, 
with the bands playing a role similar to the spin as far as the matrix structure is concerned in the former and latter cases, respectively. 
The $d+d$ pairing yields good fits to penetration depth, specific heat, and NQR data well below and closer to $T_{c}$ alike \cite{Pang2018,Smidman2018},
as discussed in the following subsection.

While $d+d$ pairing defined in the band basis provides a direct interpretation of the experimental results, its stability is more naturally 
addressed using microscopic matrix-pairing candidates defined in the orbital/spin space of the paired electrons. 
Matrix-pairing states which transform according to the irreducible representations of the point group can be constructed from the decomposition
 of the products of two-orbital, 
or more generally, spin-orbit coupled multiplets of definite symmetry. This approach was illustrated in the alkaline Fe-selenides, 
which are also strongly-correlated multi-band superconductors. 
[The properties of other Fe-selenide superconductors with a similar or higher 
$T_c$ as the alkaline 
Fe-selenides, including the Li-intercalated iron selenides \cite{Lu2015} and even 
the single-layer FeSe, the 
$T_c$ record holder of the iron-based superconductors \cite{Wang2012b}, are similar \cite{Si2016}.]
 In spite of the difference in the nature of their basic constituents, these Fe-based superconductors 
 remarkably share some of the experimental signatures
 that are similar to those in CeCu$_{2}$Si$_{2}$, namely, fully-gapped superconductivity, as indicated 
 by ARPES experiments~\cite{Mou2011, Wang2011, Xu2012, Wang2012}. 
This is supported by a spin resonance in the INS spectrum at the wave vector $\mathbf{Q}_{\text{Alkaline FeSe}}=(0.5, 0.25, 0.5)$~\cite{Park2011, Friemel2012},
which is distinct from what one could expect 
from the sign-changing $s$-wave pairing. In any case,
this latter scenario is unlikely given the absence of hole pockets at the center of the Brillouin zone.
\onlinecite{Nica2017} introduced an $s\tau_{3}$ matrix-pairing state, which consists of 
an
 $s$-wave form factor multiplied by a $\tau_{3}$ Pauli matrix in the space of the Fe $d_{xz/yz}$ orbitals. 
Because the $s\tau_{3}$ matrix does not commute with the symmetry-dictated kinetic part, the multi-orbital
$s\tau_{3}$ pairing is equivalent to $d+d$ pairing 
in the band basis~\cite{Nica2017, Nica2021}. On the other hand, $s\tau_{3}$ transforms as a single $B_{1g}$
 irreducible representation of the $D_{4h}$ point group. 
This implies that $d+d$ pairing also belongs to the same representation and that it preserves both point-group and time-reversal symmetries. 
When the normal-state band splitting near the Fermi level is 
small,
the BdG
quasiparticle spectrum shows a full gap everywhere in the Brillouin zone.  
Generically, the BdG spectrum is always nodeless everywhere on the Fermi surface.
Away from the Fermi surface, nodes can occur in the BdG spectrum when
the band splitting exceeds a certain threshold.
However, in strongly correlated systems only
nodal excitations on the Fermi surface are long lived and, correspondingly,
sharply defined; any
putative nodal excitations away from the Fermi surface
involve a large correlation-induced damping in the normal state, and the distinction between nodal and
gapped excitations is obviated \cite{Nica2017,Nica2021}. 
Finally, we note that the $s\tau_{3}$ 
and the equivalent $d+d$ pairings are energetically favored: they are stabilized in 
a multi-orbital $t-J_{1}-J_{2}$ model~\cite{Nica2017} 
in the regime where $A_{1g}$ and $B_{1g}$ pairing channels are quasi-degenerate.

Following the important precedent of the alkaline Fe-selenides, \cite{Nica2021} 
constructed a microscopic candidate for even-parity, spin-singlet $d+d$ pairing which incorporates 
the nature of the electronic states in CeCu$_{2}$Si$_{2}$. Matrix-pairing candidates can be constructed 
within the quasi-localized $f$ electron sector, corresponding to $f-f$ pairing, but also in the $f-c$ and $c-c$ sectors, 
where $c$ stands for a conduction electron. As indicated by several experiments~\cite{Goremychkin1993, Rueff2015, Amorese2020} 
and by LDA+DMFT studies~\cite{Pourovskii2014}, the $^{2}F_{5/2}$ electron states split under the influence of the crystalline-electric-field into a ground-state $\Gamma_{7}$ Kramers doublet and excited $\Gamma_{6}$ and $\Gamma_{7}$ doublets. 
Within the $f-f$ pairing sector, the product of two ground-state $\Gamma_{7}$ doublets decomposes into $\Gamma_{1}, \Gamma_{2}$ 
and $\Gamma_{5}$ irreducible representations. As previously discussed, CeCu$_{2}$Si$_{2}$ 
does not show signs of multiple superconducting transitions, implying that two-component pairing states belonging to $\Gamma_{5}$ 
are unlikely to occur. From the remaining two representations, the matrix associated with $\Gamma_{2}$ is symmetric and thus incompatible 
with the even-parity, spin-singlet nature of the pairing candidate. The only possible pairing candidate within the $f-f$ sector 
is a matrix belonging to the identity $\Gamma_{1}$ representation. Because this matrix transforms trivially under the point group,
 the symmetry of $f-f$ pairing states are determined entirely by the form factor. This implies that $f-f$ pairing is not likely to support 
 $d+d$ pairing. \onlinecite{Nica2021} considered an alternative in the $f-c$ pairing sector. 

\noindent \begin{figure}[t!]
\includegraphics[width=0.7\columnwidth]
{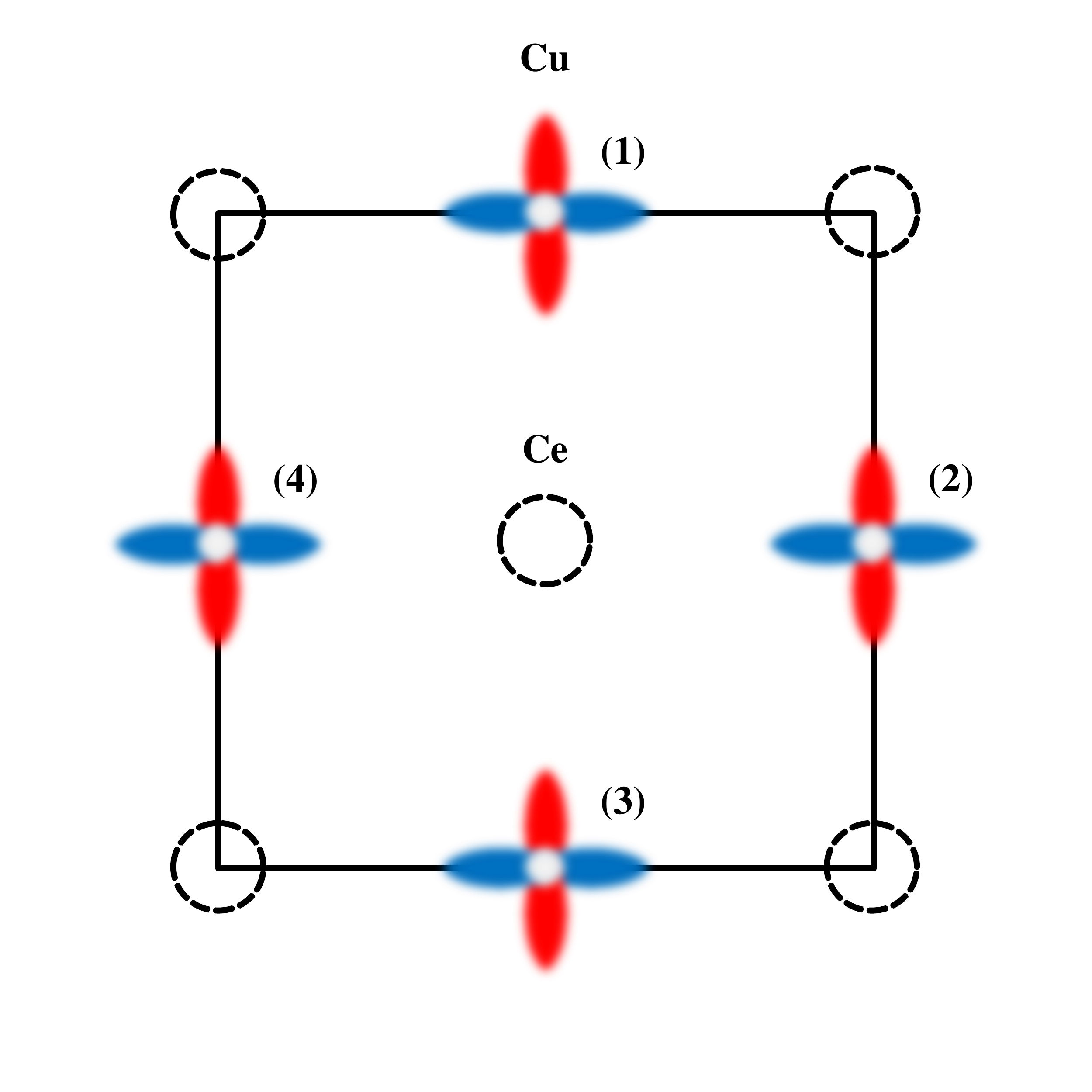}
\caption{Single Cu plane in the unit cell of CeCu$_{2}$Si$_{2}$. The four sites labeled $(1)-(4)$ correspond to Cu $d_{x^{2}-y^{2}}$ orbitals in the plane. The dashed-line circles represent the Ce sites projected onto the Cu-plane. Reproduced from \onlinecite{Nica2021}, under a Creative Commons Attribution 4.0 International License.}
\label{Gmm_6_c}
\end{figure}

\noindent Conduction electron states which belong to the $\Gamma_{6}$ irreducible representation can be constructed 
by first taking linear combinations of the Cu $d_{x^{2}-y^{2}}$ orbitals which transform as $(p_{x}, p_{y})$ within each unit cell:

\noindent \begin{align}
p_{x} = & d^{(4)}_{x^{2}-y^{2}} - 
d^{(2)}_{x^{2}-y^{2}} 
 \\
p_{y} = & d^{(1)}_{x^{2}-y^{2}} - d^{(3)}_{x^{2}-y^{2}}, 
\end{align}

\noindent as illustrated in Fig.~\ref{Gmm_6_c}. The spin-orbit coupling can be incorporated to obtain the $\Gamma_{6}$ states

\noindent \begin{align}
\Psi_{\Gamma_{6}; 1/2} = & \frac{i}{2} \left[ p_{x}+\text{i}p_{y} \right] \phi_{-1/2} \\
\Psi_{\Gamma_{6}; -1/2} = & \frac{i}{2} \left[ p_{x}-\text{i}p_{y} \right] \phi_{1/2},
\end{align}

\noindent where the $\phi$'s denote spin-1/2 states. Note that the four $d$ orbitals are localized on distinct sites in the unit cell. The $\Psi$ states are examples of a Zhang-Rice construction \cite{Zhang1988}.  The decomposition of the products of $f$-electron $\Gamma_{7}$ doublets,
 belonging to the ground-state multiplet, and $\Gamma_{6}$ conduction electron doublets includes 
 a sign-changing $\Gamma_{3}$ irreducible representation. When multiplied by a featureless $s$-wave form factor,
  the matrix associated with $\Gamma_{3}$ $f-c$ pairing is the analogue of the $s\tau_{3}$ pairing introduced in the context 
  of the alkaline Fe-selenides. $s\Gamma_{3}$ thus provides a microscopic candidate for $d+d$ pairing in CeCu$_{2}$Si$_{2}$. 
  Evidence supporting this type of pairing was provided by x-ray absorption spectroscopy experiments~\cite{Amorese2020} 
  which indicated a finite admixture of the $f$-electron $\Gamma_{6}$ in the ground-state of CeCu$_{2}$Si$_{2}$. 
It is important to recall that the microscopic candidate for $d+d$ pairing introduced in \cite{Nica2021} 
was constructed using only the point-group symmetry and a minimal input provided by the nature of the lowest-energy $4f$ Kramers doublet. 
In spite of its simplicity, this construction (i)  demonstrates how $d+d$ pairing can emerge in principle, 
and (ii) provides a well-defined microscopic candidate for any future detailed theoretical studies of the pairing symmetry 
in CeCu$_{2}$Si$_{2}$ that
also incorporate the complex band-structure of the normal state.

Sign-changing $s_{+-}$ pairing states were also advanced to explain the gapped, sign-changing superconductivity 
in CeCu$_{2}$Si$_{2}$~\cite{Li2018,Ikeda2015}. We briefly summarize two of the most important differences 
between $d+d$ and $s_{+-}$ pairing states. Firstly, although both candidates are sign-changing and therefore conducive to a 
large peak in the INS intensity inside the superconducting gap, 
they also imply very different ways of involving the states on the Fermi surface.
\onlinecite{Li2018} (see also \onlinecite{Ikeda2015}) carried out DFT+U calculations,
which capture neither the Kondo effect nor the associated renormalization towards
heavy single-electron excitations. 
Physically, the
proposed
$s_{+-}$
picture 
invokes a wave vector  that spans the distance
between the heavy cylindrical Fermi surface  (red pockets in Fig.~\ref{BZ_schm}) 
and the hole  pocket near the $Z$-point (bulk Brillouin zone) projected 
from 
light bands (not shown), which does not generate enough spin spectral weight for either 
the observed antiferromagnetic order or the observed INS spectrum in the superconducting state. A lack of such extended nesting between  these different surfaces can also be inferred experimentally from the ARPES results (Sec.~\ref{ARPESsec}) due to the electron and hole pockets being observed to have very different shapes and effective masses. In contrast, the $d+d$ pairing implies a wave vector spanning within the same cylindrical heavy Fermi surface  (red pockets in Fig.~\ref{BZ_schm}). 
The latter picture is naturally associated with a realistic heavy-fermion SDW  instability, due to the enhanced density-of-states on these pockets.
Secondly, $d+d$ and $s_{+-}$ pairing states have distinct nodal structures. As already discussed, 
the
$d+d$ pairing state has no nodes on the Fermi surface (see Fig.~\ref{BZ_schm}). 
By contrast, the $s_{+-}$ pairing state has gap zeroes that would generally be expected to intersect the extended hole Fermi surface of CeCu$_{2}$Si$_{2}$, leading to nodal excitations. This is different from the case of Fe-based superconductors, which typically have disconnected hole and electron pockets at the zone center and edges. These points imply that the $s_{+-}$ picture is not viable.

We conclude this section by briefly revisiting the effects of disorder 
on the paired states in CeCu$_{2}$Si$_{2}$. As mentioned in Sec.~\ref{impSec},
the weak suppression of $T_{c}$ in electron-irradiated samples was argued to point towards a more conventional order-parameter 
that
 does not change sign~\cite{Yamashita2017, Takenaka2017}, an interpretation 
which usually relies on 
 the perturbative
  Abrikosov-Gor'kov theory.
  However, $d$-wave pairing in strongly correlated settings is expected to be
  much less sensitive to disorder introduced via non-magnetic potential 
  scattering~\cite{Anderson1997}. Studies in
  models with strong, short-range exchange interactions 
  are consistent with this expectation~\cite{Chakraborty2017, Garg2008}. 
  This implies that $d+d$ pairing states are also robust against this type of disorder in a broader class of materials 
  with similar strong exchange interactions. These include for instance the alkaline Fe-selenides, 
  where the $d+d$ state in the form of $s\tau_{3}$ pairing was stabilized in a multi-orbital $t-J_{1}-J_{2}$ model~\cite{Nica2021}. 
We expect that strong correlations also protect the $d+d$ pairing state in CeCu$_{2}$Si$_{2}$. 
In contrast, as already mentioned above, 
$T_{c}$ can be sharply suppressed in CeCu$_{2}$Si$_{2}$ via atomic substitution, 
as is the case for instance in high-$T_{c}$ superconductors with Zn substituted for Cu on the CuO$_2$ planes \cite{Loram1990}.

\subsection{Analysis of experimental results with the $d+d$ model}

\begin{figure}[t]
	\begin{center}
		\includegraphics[width=0.99\columnwidth]{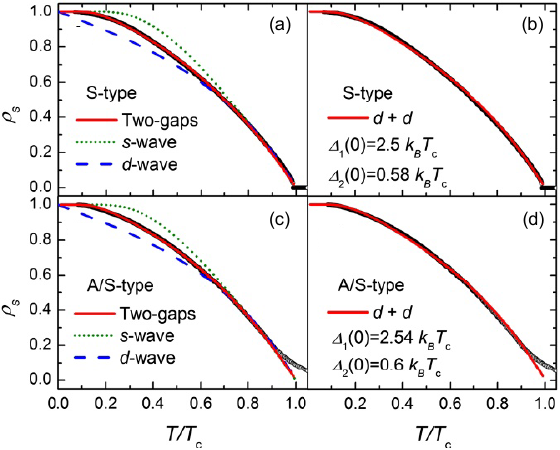}
	\end{center}
	\caption{Temperature dependence of the superfluid density derived from penetration depth measurements using the tunnel-diode-oscillator-based method from. Panels (a) and (b) display fits to the superfluid density of an $S$-type sample with an isotropic two gap model and $d+d$ band-mixing pairing model, respectively, while (c) and (d) show the corresponding results for the $A/S$-type sample. Reproduced from \onlinecite{Pang2018}.}
	\label{Fig5}
\end{figure}

Upon converting the penetration depth data measured using the tunnel-diode-oscillator-based method  to the superfluid density, Pang \textit{et al.} found that the temperature dependence could be described both by an isotropic two-gap model, as well as one for the $d+d$ band mixing pairing state \cite{Pang2018,Smidman2018}, which is displayed in Fig.~\ref{Fig5}. For the latter case, a simple model of the gap function is given by $\Delta(T,\phi)=[(\Delta_1(T){\rm cos}2\phi)^2+(\Delta_2(T){\rm sin}2\phi)^2]^{\frac{1}{2}}$, which has a four-fold oscillatory component where one of the gap parameters corresponds to the gap minimum and the other to the maximum. The basis for applying this model is explained in the previous subsection, and it is found that this  describes the data across the whole temperature range well. The fitted values of the gap parameters for measurements of the $S$-type sample were $2.5k_{\rm B}T_{\rm c}$ and $0.58k_{\rm B}T_{\rm c}$, where the small but finite gap minimum ensures a nodeless gap across the Fermi surface, and is close to the magnitude obtained from the low temperature analysis of $\Delta\lambda(T)$ (Sec.~\ref{FullgapSec}). This model can also fit the temperature dependence of the specific heat \cite{Pang2018,Smidman2018}, including the data previously reported by Kittaka \textit{et al.} \cite{Kittaka2014}. 

In the case of the recent NQR measurements, this $d+d$ band mixing pairing model can  well account for $1/T_1(T)$ across the whole temperature range, including the  deviation from $T^3$ behavior resolved at the lowest temperatures from recent measurements \cite{Kitagawa2017,Smidman2018}. Although the simple two-band BCS model can also describe the low temperature $1/T_1(T)$ results, it is less accurate at elevated temperatures, where it deviates from the data, culminating in the prediction of  a pronounced Hebel-Slichter coherence peak below $T_{\rm c}$. 
 Such an enhancement, which is a hallmark of conventional BCS superconductivity,  is absent from the data (Fig.~\ref{Fig3a}), which is in-line with a sign-changing order parameter. In the $s_{+-}$ scenario this peak is somewhat suppressed, but still present in the model. In analogy with the Fe-based superconductors, effects such as quasiparticle damping and impurity-induced bound states (in the case of $s_{+-}$ pairing) could potentially account for the deviations from these two models \cite{Bang2017}. On the other hand, for a $d+d$ band mixing pairing state the coherence peak is naturally avoided due to the sign change of the intraband pairing component \cite{Kitagawa2017,Smidman2018}.

\section{Perspectives}
\label{PersSec}

Despite the progress  made on this prototypical heavy-fermion superconductor, 
a number of points are worthy of further investigations.

 Although the band-mixing $d+d$-pairing state can account for all the experimental results, more direct experimental evidence 
 for such a scenario is still lacking. To unambiguously discriminate between different fully gapped models likely requires 
 high-resolution momentum resolved experimental probes of the superconducting gap at very low temperatures, 
 which is very challenging. 
 In addition, while recent proposals have given a microscopic basis to the $d+d$-pairing state
  \cite{Nica2021}, 
 a fully developed microscopic theory
 for CeCu$_2$Si$_2$ is still necessary.
Indeed, developing fully microscopic theories for strongly correlated superconductors remains a grand challenge of condensed matter physics.
 
The effect of nonmagnetic potential scattering on CeCu$_2$Si$_2$ still lacks a complete theoretical and experimental understanding. This is especially so concerning the variable sensitivity of the superconductivity to  substitutional disorder, which appears to be site-dependent, as well as to various types of lattice rearrangement, such as that induced by electron irradiation.  There is a pronounced size dependence for substitutions on the Ce-site, where the magnitude of the $T_c$-depression is found to be \textit{anticorrelated}   to the volume of the so obtained ``Kondo hole'', while Ge atoms exchanged for Si are less strong pairbreakers. The origin of this nonuniversal  impact of substitutional disorder on the superconductivity of CeCu$_2$Si$_2$, as well as the dichotomy between ``harmful" and ``harmless" (e.g., electron-irradiation-induced) disorder are interesting open questions to be unraveled by future work.

 We also note that CeCu$_2$Si$_2$ has often been regarded as a  prototypical example of both heavy-fermion superconductivity, as well as SDW-type quantum criticality, but the extent to which the findings extend to other heavy-fermion systems is currently unclear. In particular, the nodeless superconducting gap structure of CeCu$_2$Si$_2$ is distinct from the clearly evidenced nodal $d_{x^2-y^2}$ superconductivity in Ce(Co,Ir)In$_5$ 
 \cite{Izawa2001,An2010,Park2008,Kasahara2008,Lu2012,Allan2013,Zhou2013}.

 The spin-excitation spectrum of CeCu$_2$Si$_2$ consists of both long-wavelength SDW-type fluctuations (paramagnons), as well as high-frequency Mott-like fluctuations of $4f$ electron spins. It is of special interest to understand the role played by these different types of spin fluctuations in either promoting or breaking apart the Cooper pairs. In this context, it is interesting that the $B_{\rm c2}(p)$ curve in the $T=0$ plane of Fig.~\ref{Fig3}(a) exhibits its maximum at a pressure which is only about half  the value of the critical pressure $p_c$ at $B=0$. When increasing the pressure at $p\ll p_c$ far from this QCP, in the absence of quantum-critical SDW fluctuations, $B_{\rm c2}(p)$ is found to increase which apparently means that superconductivity becomes strengthened. 
However, when further approaching the QCP (at $p_c/2 < p < p_c$), under increasingly dominant SDW-type quantum critical fluctuations, 
$B_{\rm c2}(p)$ turns out to decrease and superconductivity deteriorates. A similar conclusion can be drawn from the evolution of $T_{\rm c}(p)$ for the low-pressure dome displayed in Fig.~\ref{twodome} and may also apply to other correlated metals showing a superconducting dome centered at an SDW-or putative SDW- type QCP, such as Ba(Fe$_{1-x}$Co$_{x}$)$_2$As$_2$ \cite{Chu2009} or CePd$_2$Si$_2$ \cite{Mathur1998}. This non-monotonic evolution suggests that the Mott-type critical excitations are pair-promoting, while the ultra-low-temperature (below $T^*=1$~K) SDW-type critical excitations in CeCu$_2$Si$_2$ are pair-breaking. Interestingly, the theoretical work of \onlinecite{Hu2021}, for an SDW-type quantum criticality of Kondo-lattice systems, 
reached a similar conclusion that the Mott-type quantum critical fluctuations at energies above $T^*$
are primarily instrumental for the Cooper-pair formation. Together, these considerations suggest that the heavy-fermion superconductivity in CeCu$_2$Si$_2$ should be compared with those of systems with local (Kondo-destroying) rather than itinerant (SDW-type) QCPs \cite{Shishido2005,Park2006,Schuberth2016,Nguyen2021,Schuberth2022,Shan2022}.

\section{Summary}
CeCu$_2$Si$_2$ was originally considered a prototypical intermediate-valence metal \cite{Sales1976}. The discovery of heavy-fermion behavior \cite{Steglich1979} in this compound led to the notion that it belongs to the family of Ce-based Kondo-lattice systems \cite{Bredl1978,Bredl1984} and, most importantly, CeCu$_2$Si$_2$ is the first discovered unconventional superconductor \cite{Steglich1979}. Over 40 years of intense research on this system have posed several severe challenges and surprising solutions most of which are  covered in this article. In the following, we briefly summarize our current knowledge on CeCu$_2$Si$_2$.

Its Kondo-lattice ground state, implying a local $J = 5/2$ spin-orbit split Hund’s rule multiplet of trivalent Ce, which is further split by the tetragonal crystalline-electric field into two $\Gamma_7$ and a $\Gamma_6$ Kramers doublets \cite{Amorese2020}, could be recently verified by ARPES experiments performed at $10$\,K \cite{Wu2021}, well below the lattice Kondo temperature of $15$\,K. These investigations revealed a `large (renormalized) Fermi surface' to which the Ce-$4f$ electrons substantially contribute, i.e., a heavy electron band near the $X$-point of the bulk Brillouin zone. For this heavy band the effective charge-carrier mass $m^*$ estimated from ARPES of $m^* \approx 120\,m_{\rm e}$ is in very good agreement with that obtained from specific-heat results at the same temperature (Fig.~\ref{Fig1new}(a)) \cite{Steglich1990}. In addition, ARPES revealed a hole band with small, but significant $4f$-contribution near the bulk $Z$-point which corresponds to the distinct Fermi surface  pocket  with moderately enhanced $m^*$($\approx 5\,m_{\rm e}$) that had been detected by magnetic quantum oscillation measurements \cite{Hunt1990,Tayama2003}. In contrast to the aforementioned ground-state and thermodynamic properties which probe the large Fermi surface  of the Kondo-lattice state of CeCu$_2$Si$_2$  at finite temperatures, transport measurements \cite{Sun2013,Shan2022} appear to be dominated down to very low temperatures by the fundamental local scattering process underlying the Kondo screening, i.e., scattering of ordinary conduction electrons from the Ce-derived localized $4f$-spins, see also \cite{Coleman1985}. Upon volume compression, Ce-based Kondo-lattice systems commonly show a strengthening of the Kondo interaction and eventually a transition into an intermediate-valence state. This has been observed for CeCu$_2$Si$_2$ as well \cite{Yuan2003,Holmes2004,Yuan2006}.

One of the characteristics of these types of materials is their closeness to magnetism. Many of them exhibit a magnetically ordered low-temperature phase in the vicinity of a QCP. While the discovery of superconductivity in CeCu$_2$Si$_2$ with a finite magnetic moment in each unit cell came as a big surprise for most researchers in the field of superconductivity, this might indeed have been expected for researchers working on superfluid $^3$He \cite{Vollhardt1990}. With the discovery of a heavy-fermion low-temperature phase in CeAl$_3$ \cite{Andres1975}, which resembles the renormalized normal phase of (charge-neutral) liquid $^3$He at sufficiently low temperatures, the question might have arisen: Is there a superconducting analogue in a heavy-fermion metal like CeAl$_3$ to the superfluid phases in $^3$He? Not surprisingly, magnetically-driven superconductivity in heavy-fermion metals was proposed quite early by theorists \cite{Anderson1984, Miyake1986, Scalapino1986} and was then gradually verified experimentally \cite{Aeppli1989, Sato2001}. In  the case of CeCu$_2$Si$_2$, it became clear from the outset that a BCS-type phonon-mediated Cooper-pairing mechanism is incapable of explaining why the non-magnetic analogue compound LaCu$_2$Si$_2$ is not a superconductor \cite{Steglich1979} as well as the drastic pair-breaking effect of certain non-magnetic impurities, notably when substituted for Cu in CeCu$_2$Si$_2$ \cite{Spille1983}. 

In more recent years, CeCu$_2$Si$_2$, along with  CeCu$_{6-x}$Au$_x$, YbRh$_2$Si$_2$ and CeRhIn$_5$, have played a prominent 
role in the understanding of heavy-fermion quantum criticality \cite{Gegenwart2008}. Theoretical studies of Kondo-lattice models have led to the notion of Kondo destruction \cite{Si01.1,Col01.1}, which characterizes Mott-type quantum criticality for an electron localization-delocalization transition. More recently, it has been argued that partial Mott quantum criticality also forms the basis for the ferromagnetic instabilities in the heavy-fermion metals YbNi$_4$(P$_{1-x}$As$_x$)$_2$ \cite{Steppke2013} and CeRh$_6$Ge$_4$ \cite{Shen2020}. In CeCu$_2$Si$_2$, it has been suggested that SDW-type critical excitations operate below an energy scale  $T^*$ that is nonzero but 
much smaller than the Kondo temperature, while the Mott-type critical excitations describe the quantum criticality above this energy scale
\cite{Gegenwart2008,Smidman2018}. Theoretical studies that incorporate the Kondo destruction physics in
 quantum-criticality-driven superconductivity have recently been developed
\cite{Hu2021}.

In the low-temperature normal state of $S$-type CeCu$_2$Si$_2$, the critical exponent of the power-law $T$-dependence of the resistivity turned out to be ambiguous, i.e. 1.5 \cite{Gegenwart1998} or 1 \cite{Yuan2003,Yuan2006}, presumably due to the spatial distribution of a magnetically ordered minority phase \cite{Stockert2011} that may modifiy the volume-integrated response in resistivity experiments. From the temperature dependences  of both $C(T)/T$ and the damping rate measured in the INS spectrum \cite{Smidman2018,Arndt2011,Gegenwart2008}, $T^*\sim1-2$~K can be inferred, which is of the same order of magnitude as the spin excitation gap in the magnetic response in the superconducting state.
Nevertheless, the linear paramagnon dispersion relation observed above the spin-gap energy $\hbar\omega_{\rm gap}$ extends to about 1.5~meV \cite{Song2021}. Except for these paramagnon excitations, the magnetic INS response comprises of Mott-type fluctuations 
of local Ce-moments with frequencies in the range $k_BT^*/\hbar$ to $k_BT_K/\hbar$. The existence of a nonzero Kondo-destruction  energy scale $k_BT^*$ that is small compared to the Kondo temperature has also been inferred from
the large kinetic energy loss as CeCu$_2$Si$_2$ goes from the normal to the superconducting state; this kinetic energy loss overcompensates the majority of the exchange energy saving in the same process \cite{Stockert2011}. This overcompensation results in a pair-formation energy that is smaller than the exchange energy by a factor of about twenty -  characteristic of magnetically-driven Cooper pairing of slowly propagating Kondo singlets \cite{Stockert2011}. As far as the magnetism in CeCu$_2$Si$_2$  is concerned, the nature of the high-field B-phase \cite{Bruls1994}, and of its QCP at about 17 T \cite{Weickert2018} as well as the first-order phase transition between this B-phase and the adjacent low-field SDW A-phase \cite{Tayama2003} need further detailed exploration. Meanwhile the field dependence of the specific heat in the superconducting state exhibits an unusual upturn at intermediate fields culminating in a strongly enhanced value just below $B_{c2}$ \cite{Kittaka2014,Kittaka2016}. The origin of this behavior still needs to be determined, especially whether it is related to a spatially modulated superconductivity \cite{Kitagawa2018} or other inferred effects of strong Pauli-paramagnetic limiting \cite{Campillo2021}.

Another striking phenomenon is the occurrence of a second superconducting dome in CeCu$_2$Si$_2$ at  pressures well above the critical pressure at which SDW order disappears. There,  the $T_c$ is around three times larger than in low pressure conditions \cite{Yuan2003,Yuan2006}. Although such a scenario was hinted at by the unusual shape of the $T_c$ vs $p$ plateau, the existence of a distinct second high pressure dome was only apparent upon doping with Ge to weaken the superconductivity (Fig.~\ref{twodome}), and suggests a different unconventional pairing mechanism at higher pressures, namely one related to valence fluctuations \cite{Yuan2003,Holmes2004}.

For a long time, CeCu$_2$Si$_2$ was believed to be a single-band $d$-wave superconductor with line nodes in the energy gap. The strongest evidence for this conclusion came from NQR measurements down to 0.1~K, which revealed the absence of a Hebel-Slichter peak at $T_c$ and a $T^3$ dependence of $1/T_1(T)$ \cite{Ishida1999,Fujiwara2008}. A $d_{x^2-y^2}$ state was concluded from INS \cite{Eremin2008,Stockert2011}, while  $d_{xy}$ was deduced from the anisotropy of the  upper critical field determined from the resistivity \cite{Vieyra2011}. This understanding was overturned by the results of low-temperature specific-heat \cite{Kittaka2014,Kittaka2016}, penetration depth \cite{Yamashita2017,Takenaka2017,Pang2018}, thermal conductivity \cite{Yamashita2017}, and more recent NQR measurements on  CeCu$_2$Si$_2$ single crystals \cite{Kitagawa2017}  which reveal a small, but finite fully open superconducting gap. Theoretical proposals to account for these findings include both isotropic (non-sign-changing) \cite{Takenaka2017,Yamashita2017} and anisotropic (sign-changing) $s$-wave pairings  \cite{Ikeda2015,Li2018}, as well as a $d+d$ matrix pairing state \cite{Nica2017,Pang2018,Nica2021}(see Table~\ref{table:table2}). 

The aforementioned $s$-wave  pairings are disfavored for the following reasons: 

(i) As discussed in Sec.~\ref{SpinDynSec}, a pronounced maximum is observed in the INS intensity inside the superconducting gap, exactly at the SDW ordering wave vector $\mathbf{Q_{\rm AF}}$. The latter equals the nesting vector $\boldsymbol\tau$ inside the warped part of the cylindrical heavy-electron band at the $X$-point of the bulk Brillouin zone \cite{Smidman2018,Wu2021}. This maximum demonstrates a sign-change of the superconducting order parameter along $\boldsymbol\tau$, which means intraband pairing as already discussed. No such sign change is possible for isotropic BCS-type pairing. In addition, such onsite pairing is unfavorable  in a heavy-fermion superconductor as the heavy charge carriers forming the Cooper pairs only have a tiny kinetic energy of order $k_{\rm B}T_{\rm K}$, which is of the same order as their renormalized Coulomb repulsion. For an onsite pairing to operate in a BCS superconductor, the kinetic energy must be much larger than the effective Coulomb repulsion. In an innovative approach, \cite{Tazai2018,Tazai2019} succeeded in showing that both phonon-mediated and electronically-driven $s$-wave heavy fermion superconductivity can arise from  higher multipole charge fluctuations. However, the \textit{magnetically-driven} nature of the superconductivity in CeCu$_2$Si$_2$ \cite{Stockert2011} necessitates sign-changing superconductivity \cite{Scalapino2012}.  More generally, such an $s$-wave pairing without  a sign change is difficult to be reconciled with the exclusion of onsite pairing  associated with the strong Coulomb repulsion of the $4f$ electrons. 

(ii) Anisotropic $s$-wave pairing also cannot explain the superconductivity of CeCu$_2$Si$_2$. In order to account for the pronounced peak observed in INS at $\mathbf{Q_{\rm AF}}$ inside the superconducting gap, there would need to be \textit{interband} nesting connected by the SDW ordering wave vector, whereas ARPES measurements \cite{Wu2021} and calculations of the renormalized electronic structure \cite{Zwicknagl1992,Zwicknagl1993}  demonstrate that this ordering wave vector must connect regions within the heavy electron pocket,  as indeed revealed by neutron diffraction \cite{Stockert2004}, see Fig.\ref{Fig3_1}(b). This confirms that there is a sign change of the order parameter within this band, in contrast to the $s_{+-}$ scenario where the sign changes between the hole and electron pockets \cite{Li2018} (see also \onlinecite{Ikeda2015}).

A $d+d$ pairing state with intra and interband components provides a natural resolution to all currently available experimental results, and is in line with the importance of the non-perturbative effect of the strong Coulomb repulsion of the $4f$-electrons in the form of a Kondo effect. The intraband $d$-wave component accounts for the sign change on the heavy warped cylindrical bands. The two distinct components added in quadrature also ensure a fully gapped Fermi surface. This pairing state belongs to a single irreducible representation of the point group, which coincides with that of a single-band $d$-wave,  and therefore implies a single transition to the superconducting phase, as observed in CeCu$_{2}$Si$_{2}$. On the microscopic level, $d+d$ pairing is equivalent to a matrix-pairing state between $f$-electrons in $\Gamma_{7}$ doublets and conduction electrons belonging to $\Gamma_{6}$ doublets. The non-trivial matrix structure ensures the presence of the two $d$-wave components in the band basis. Similar $d+d$ candidates were proposed in the context of the alkaline Fe-selenides~\cite{Nica2017, Nica2021}, suggesting a common theme in unconventional superconductivity. Nevertheless, in line with other classes of unconventional superconductors, the unambiguous determination of the pairing state and mechanisms of CeCu$_2$Si$_2$ still requires a  fully developed microscopic theory together with additional experimental results able to discriminate between different scenarios.

Taking all these together, CeCu$_2$Si$_2$, the very first unconventional superconductor ever discovered, continues to grow in its role as a 
model system for strong correlation physics. The historical intuition about CeCu$_2$Si$_2$ as a solid-state generalization
of the superfluidity observed in liquid $^3$He inspired the early considerations regarding the interplay between antiferromagnetic
correlations and $d$-wave superconductivity. The observation that the Cooper pairs in CeCu$_2$Si$_2$ are formed by the extremely heavy charge carriers existing in the low-temperature phase of the Kondo lattice proved the superconducting pairing mechanism to be incompatible with the conventional one of BCS theory. In modern times, CeCu$_2$Si$_2$, like CeCu$_{6-x}$Au$_x$ \cite{Lohneysen1994,Schroder2000}, CePd$_2$Si$_2$ \cite{Mathur1998}, CeCoIn$_5$ \cite{Paglione2003}, and CeRhIn$_5$ \cite{Shishido2005,Park2006}, has served as a model system for heavy fermion antiferromagnetic quantum criticality.
Intriguingly,
here, the Landau-type SDW-type quantum criticality interplays with the beyond-Landau Mott-type quantum criticality in different 
energy ranges below the Kondo temperature. Moving further 
to just over the 
past few years, CeCu$_2$Si$_2$ has emerged as a model system for multiband superconductivity
with strongly correlated carriers. We certainly won't be surprised if the future will bring out yet more surprises about 
the superconductivity in CeCu$_2$Si$_2$ and related heavy-fermion systems.

\subsection*{Acknowledgements}  

We would like to thank Wolf A{\ss}mus, Ang Cai, Lei Chen, Piers Coleman, Pengcheng Dai, Onur Erten, Zach Fisk, Jacques Flouquet, Philipp Gegenwart, Christoph Geibel, Norbert Grewe, Malte Grosche, Haoyu Hu, Kenji Ishida, Kevin Ingersent, Lin Jiao, Hirale S. Jeevan,  Stefan Kirchner, Michael Lang, Michael Loewenhaupt,  Alois Loidl, Bruno L\"{u}thi, Brian Maple, Kazumasa Miyake, Guiming Pang,
 Silke Paschen, Jed H. Pixley, Noriaki Sato, Doug Scalapino, Erwin Schuberth, Andrea Severing, Yu Song, G\"{u}nter Sparn, Greg R. Stewart,  Joe D. Thompson, Hao Tjeng, Roxanne Tutchton, Hilbert von L\"{o}hneysen,   Franziska Weickert,
 Steffen Wirth, Zhongzheng Wu, Rong Yu, Jinglei Zhang, Jian-Xin Zhu,  and
Gertrud Zwicknagl for useful discussions. 
Part of these discussions took place at the
2019 Zhejiang Workshop on Correlated Matter, which provided the initial motivation for this work.
This work was supported by the National Key R$\&$D Program of China (2022YFA1402200), the Key R$\&$D Program of Zhejiang Province, China (2021C01002), the National Natural Science Foundation of China (12222410, 12034017, 11974306 and 12174331), the Zhejiang Provincial Natural Science Foundation of China (LR22A040002), at Arizona State University by the NSF Grant No. DMR-2220603 and an ASU startup grant; at Rice University by the NSF Grant No. DMR-2220603 and the Robert A. Welch Foundation Grant No. C-1411. Q.S.\ acknowledges 
 the hospitality of the Aspen
Center for Physics, which
is supported by NSF Grant No. PHY-2210452.

%

\end{document}